\documentclass[12pt,a4paper]{article}
\usepackage{amssymb}

\usepackage{graphicx}
\usepackage{amsmath}

\input{tcilatex}

\begin{document}

\title{Statistical properties of an ensemble of vortices interacting with a
turbulent field}
\author{Florin Spineanu and Madalina Vlad \\
Association Euratom-MEC Romania, NILPRP \\
MG-36, Magurele, Bucharest, Romania\\
\emph{and} \\
Research Institute for Applied Mechanics \\
Kyushu University, Kasuga 816-8580, Japan}
\maketitle

\begin{abstract}
We develop an analytical formalism to determine the statistical properties
of a system consisting of an ensemble of vortices with random position in
plane interacting with a turbulent field. We calculate the generating
functional by path-integral methods. The function space is the statistical
ensemble composed of two parts, the first one representing the vortices
influenced by the turbulence and the second one the turbulent field
scattered by the randomly placed vortices.
\end{abstract}

\tableofcontents

\section{Introduction}

This work presents a study of the statistical properties of a system of
vortices interacting with random waves. This is motivated by the necessity
to describe quantitatively the statistics of a turbulent plasma in the
regime where structures are generated at random, persist for a certain time
and are destroyed by perturbations. This regime is supposed to be reached at
stationarity in drift wave turbulence when there is a competition between
different space scales. These scales range from the radially extended eddies
of the ion temperature gradient (ITG) driven modes to the intermediate $%
\sqrt{L_{n}\rho _{s}}$ interval where the scalar nonlinearity is dominant
over the vectorial one and with random fast decays to few $\rho _{s}$ scales
where robust vortex-like structures are generated. In a physical model the
description of these states requires to consider simultaneously the long
time response and the much faster events of trapping of the energy in small
vortices, condensed from transient Kelvin-Helmholtz instabilities of the
locally ordered sheared flow.

The analytical model we propose here is essentially \emph{nonperturbative}
in the sense that the structures are explicitely represented and we take
into account their particular analytical expression (or a reasonable
approximation). It is in the spirit of the semiclassical methods, frequently
used in field theory. In this sense this treatment presents substantial
differences compared with the more usual approaches, aiming in general to
obtain a renormalization of the linear response or to calculate the
correlations of the fluctuating field using a closure method. When
structures are present in the plasma any perturbative approach is
inefficient, because it can only go not too far beyond Gaussian statistics, 
\emph{i.e.} calculate few cumulants beyond the second one. The problem
consists in that they take as the base-point for developing perturbative
expansions the state of equilibrium of the plasma or a Gaussian ensemble of
waves. Or, from this point it is impossible to reach the strongly correlated
states of coherent structures. The idea of semiclassical methods is
precisely to place the equilibrium state on the structures and then to
explore a neighborhood in the space of system's configurations, to include
random waves. This is the approach we will adopt in the present work. To
have a tractable problem we assume the simplest case where the vortices are
not created or destroyed dynamically. The final outcome from this approach
is a list in which a particular dependence of a contribution to the
correlation, \emph{e.g.} the exponent $\mu $ in $\left\langle \phi \phi
\right\rangle _{\mathbf{k}}\sim k^{-\mu }$, is associated to a particular
contribution in the physical model: statistics of the gas of vortices,
interaction energy, nonlinearity of the physical model, etc.

\subsection{The formulation of the problem}

In agreement with the experimental observation and numerical simulations 
\cite{KraichnanMontgomery} it has been found theoretically that the
nonlinear differential equations describing strongly nonlinear electrostatic
drift waves in two-dimensional plasma can have (1) turbulent solutions,
consisting of very irregular fluctuations which can only be described by
statistical quantities (irreducible correlations = cumulants); and (2)
solutions that are cuasi-coherent structures of vortex type, which are
remarkably robust and for which we have in certain cases explicit analytical
expression. Although there is no conceptual difference between these two
aspects, we can say that we have two manifestations of the nonlinearity: one
is nonlinear mode coupling and energy transfer between waves ; and the
second is the generation of organized flow with vortical pattern with strong
stability and coherence of form. A first approximation is to break up the
field (the electric potential $\phi $) in two distinct elements : vortices
and random waves. We will suppose that there is a finite number $N$ of
vortices in a two-dimensional plasma and that these vortices have random
position. The vortices are individually affected by the turbulent
background. The turbulent background, in turn, is affected by the presence
of the structures at random positions. In addition the turbulent background
has statistical properties generated by the nonlinear nature of the waves
interactions, even at amplitudes below those necessary to condense vortices.
Random growth and decay of modes at marginal stability is included as a
drive with Gaussian statistics.

Consider the field $\phi ^{Vortices}\equiv \phi _{V}$ of the discrete set of 
$N$ vortices individually represented by the electric potential $\phi
_{s}^{\left( a\right) }\left( x,y\right) $ 
\begin{equation}
\phi _{V}\left( x,y\right) =\sum_{a=1}^{N}\phi _{s}^{\left( a\right) }\left(
x,y\right)  \label{phiv}
\end{equation}
interacting with a turbulent wave field $\phi ^{waves}\equiv \phi $. The
total field is 
\begin{equation}
\varphi \left( x,y\right) =\sum_{a=1}^{N}\phi _{s}^{\left( a\right) }\left(
x,y\right) +\phi \left( x,y\right)  \label{fitot}
\end{equation}
and we want to determine the statistical properties of the field $\varphi
\left( x,y\right) $. We will construct an action functional and we will
calculate the generating functional of the irreducible correlations
(cumulants) of the fluctuating field $\varphi $. The action functional is
expressed in terms of the field $\varphi $ which we see as composed from
vortices and turbulence, Eq.(\ref{fitot}). For example, the two-point
correlation is composed of four terms 
\begin{eqnarray}
\left\langle \varphi \varphi \right\rangle &=&\left\langle \phi
^{Vortices}\phi ^{Vortices}\right\rangle +\left\langle \phi ^{Vortices}\phi
^{waves}\right\rangle  \notag \\
&&+\left\langle \phi ^{waves}\phi ^{Vortices}\right\rangle +\left\langle
\phi ^{waves}\phi ^{waves}\right\rangle  \label{eq3}
\end{eqnarray}
Our procedure consists in absorbing the two intermediate terms into the
first and the last terms. This is done by calculating the auto-correlations
of each component taking into account the presence of the other. In this
operation the second and the third terms, although are identical in Eq.(\ref
{eq3}) are regarded differently: the second term is seen as the contribution
of the turbulent background to the auto-correlation of a gas of vortices and
the third term is seen as the contribution of a set of vortices to the
auto-correlations of a turbulent field.

Therefore we consider that the action functional is composed of two distinct
parts: $S_{V\varphi }\equiv $ the action for the system of vortices
interacting with the random field; and $S_{\varphi V}\equiv $ the action for
the random field interacting with the vortices. \ Then the statistical
ensemble of realizations of the fluctuating field $\varphi \left( x,y\right) 
$ will actually consists of the Cartesian product of two distinct parts. The
generating functional will be 
\begin{equation}
Z=Z_{V\varphi }Z_{\varphi V}  \label{eq6}
\end{equation}
where each factor is calculated using the action defined above. Although the
generating functional is factorized, which corresponds to splitting the
statistical ensemble into two parts, these two parts are not independent
since each will be calculated such as to include the effect of the field
from the other subsystem.

The interaction with an external current must be included in each of the
action in order to calculate the correlations using functional derivatives.
In the full action the current is introduced by adding a term in the
Lagrangian density, $J\varphi $, where $\varphi $ is the total field,
vortices plus random waves. When we separate the two components we have $%
\varphi =\phi ^{Vortices}+\phi ^{waves}$ and $J\varphi =J\phi
^{Vortices}+J\phi ^{waves}$ . This expression must be inserted in the action
functional for the full system, before the intermediate terms in Eq.(\ref
{eq3}) are absorbed into the first and the last. This means that the \emph{%
same} current $J$ will appear in the final expressions of the two generating
functionals Eq.(\ref{eq6}) 
\begin{equation}
Z\left[ J\right] =Z_{V\varphi }\left[ J\right] Z_{\varphi V}\left[ J\right]
\label{eq2044}
\end{equation}
It will be shown below that the value of the field at a particular point can
be obtained by functional derivation to the current at that point $\varphi =%
\frac{\delta }{\delta J}$. We obtain the average of the fluctuating field as 
\begin{eqnarray}
&&\left\langle \phi \right\rangle =\left. \frac{1}{Z\left[ J=0\right] }\frac{%
\delta }{\delta J}Z\left[ J\right] \right| _{J=0}  \notag \\
\hspace*{-1cm} &=&\left. \frac{1}{Z_{V\varphi }\left[ J=0\right] }\frac{%
\delta }{\delta J}Z_{V\varphi }\left[ J\right] \right| _{J=0}+\left. \frac{1%
}{Z_{\varphi V}\left[ J=0\right] }\frac{\delta }{\delta J}Z_{\varphi V}\left[
J\right] \right| _{J=0}  \label{eq2049}
\end{eqnarray}
The two-point correlation is obtained from the generating functional by
applying two times this functional derivative, with the current in two
distinct points and taking finally the current to zero.

We now explain how this will be done effectively.

\subsection{Outline of the present analytical approach}

\subsubsection{Gas of vortices in the turbulent background}

In the case of the vortices interacting with random waves, the basic
reasoning takes into consideration two distinct elements.

The first is concerned with the statistical properties of a collection of
discrete vortices with zero, or very short, range of interaction. At this
stage of the problem the particular shape of the potential distribution in a
vortex is not essential and will be simplified in those situations where
only the positions of the centers are important. The statistical ensemble
consists of the configurations of randomly placed vortices, seen as a dilute
gas.

The second aspect is concerned with the interaction between one (generic)
vortex and the random waves in the surrounding turbulence and calculates the
effect on the form of the potential of the vortex. Instead of the exact
solution we will have now a form resulting from the scattering due to the
random perturbation produced by the turbulent background.

The statistical properties result from the combination of the two elements,
which will now be explained briefly.

\bigskip

The first problem is very similar to the Coulomb gas in two dimensions or to
any problem related to dilute gas of interacting particles. The partition
function is calculated using the sum of the energy of the individual
vortices and the energy of interaction. The first part can be calculated
from the exact analytical solution of the differential equation of the model
(Eq.(\ref{solMH}) below). The second part contains the sum of the energies
of pair interaction, a problem that is principially complicated in our case, 
\emph{i.e.} in turbulent plasma.\ Our \emph{real} vortices are neutral (they
are simply a deformation, stable and regular, of the scalar potential of the
velocity) and there should be no interaction of Coulomb or Ampere type.
However there is an elastic medium between the vortices and any motion of
one of them generates sound waves that may couple with the potential of
other vortices, influencing their motion. This may suggest that the
interaction is of the same nature as the Casimir effect but this is beyond
our primary interest here. What we minimally need to include in our
partition function is the effect of scattering of the vortices at close
encounter, since we do not take into account in the present treatment the
variation of the number of vortices due to merging (which would imply a
chemical potential). There is an intrinsic spatial scale in all aspects
related with drift-wave vortices and this is the sonic Larmor radius,
altered by the combined effect of the diamagnetic and a uniform flow. We
show below that the pair interaction contains a kernel with fast spatial
decay, $K_{0}\left( r/\rho _{s}\right) $ (the modified Bessel function).
There are several reasons to assume that this is a physically correct choice
(see \cite{FlorinMadi5}).

\bigskip

The second aspect of the problem of vortices acted upon by turbulence is
related to the direct modification of the exact vortex by the random waves
around it. The case of a single vortex interacting with random waves has
been treated in the papers \cite{FlorinMadi1}, \cite{FlorinMadi2}. The
starting point is the exact vortex-type solution of the differential
equation we investigate. This is of course a non-random object and when it
is isolated the statistical ensemble is trivial. However its interaction
with random waves of the background turbulence makes it also a fluctuating
object. We calculate the statistical properties of the ensemble of states of
the fluctuating potential corresponding to the shapes of the vortex-type
solution.

The action functional is extremum at the vortex solution, whose explicit
expression we will use. A purely turbulent field also realizes the extremum
of the action, but our separation between structures and random waves
amounts to an approximative representation of the turbulent field, as a
complimentary part to the structures. All configurations consisting of
structures with randomly deformed, fluctuating shapes, together with random
waves are a very good approximation of reality and must be found in close
proximity of the extremum of the action. Then we have to explore the
functional space of the system's configuration in the neighborhood of the
structure (the extremum) trying to include as many as possible nearby
configurations in the calculation of the generating functional. This will
include into the generating functional the turbulent field, besides the
structure. In few words, the idea is that the structure and the random
waves, although very different in geometry, share a common property: they
obtain or are very close to the extremum of the action functional. The
technical procedure will consists of expanding the action functional to
second order around the structure and to integrate over the space of
configurations. This will automatically exclude bad approximations, \emph{%
i.e.} the states which are too far from realizing the extremum of the action
since for them the Boltzmann weight is exponentially small.

We conclude by noting that this is the standard semiclassical treatment \cite
{QCDinstantons}. In this sense it is similar to the treatment of vortex
statistics of the Abelian-Higgs model of superfluids \cite{Schaposnik} and
of many other systems.

\subsubsection{The turbulent field influenced by random vortices}

For the turbulent field interacting with vortices, our approach combines two
distinct elements. The first is the inclusion of the vortices as a random
perturbation in the generating functional of the turbulent waves. The second
is a perturbative treatment of the intrinsic nonlinearity of the turbulent
waves, already modified by the inclusion of the effect of the random
vortices.

In the first part we follow the similar approaches as for the electron
conduction in the presence of random impurities, or as for flexible polymers
in porous media. For the case where the vortices have uncorrelated random
positions and take at random positive or negative amplitudes of equal
magnitude we find that the problem is mapped onto the \emph{sine}-Gordon
model (actually \emph{sinh}). For our model equation this will amount to a
renormalization of the coefficient which plays the role of a physical
``mass'' of the turbulent field \cite{pf25-82-1838}, \cite{Su}, or, in other
terms to a shift of the spatial scale from $\rho _{s}\left( 1-v_{d}/u\right)
^{-1/2}$ to higher values.

The second part consists of a systematic perturbative treatment of the
nonlinearity in order to get, as much as possible, a correct representation
of the nonlinear content of the field of the turbulent waves (this means the
nonlinear interaction and nonlinear energy transfer between low amplitude
random waves \cite{HM}, \cite{Horton1}, \cite{Balescu}, \cite{Diamond}). The
nonlinearity is included in a functional perturbative treatment where the
turbulent plasma is driven by random rise and decay of modes at marginal
stability. This induces a diffusive behavior of the turbulent field, at
lowest order. We develop the treatment to one-loop, which means of order two
in the strength of the nonlinearity.

One may inquire if this is not in contradiction with the separation operated
at the beginning, where the most characteristic aspect of the nonlinearity,
the generation of structure, is treated separately. However, it is clear
that there is no danger of overlapping and double counting of the nonlinear
effects. Since a statistical perturbative treatment is essentially an
expansion in a parameter representing the departure from Gaussian statistics
we can only hope to include higher cumulants beyond the second one (which
means Gaussianity) but only few of them are accessible to effective
calculation. Or, any structure needs a very large number of cumulants since,
by definition, is an almost coherent field. It is illusory to try to capture
the structure using a perturbative treatment starting with a state of
equilibrium (\emph{i.e.} no perturbation or, alternatively, a Gaussian
collection of linear waves). Since any term of the perturbation can be
represented by a Feynman diagram, we face the well known problem that the
proliferation of diagrams at high orders leads to an effectively intractable
problem. This is actually one reason for the use of the semiclassical
methods.

\section{The physical model}

\subsection{The equation}

The model of the ion drift instability in magnetically confined plasmas can
be formulated using the fluid equations of continuity and momentum
conservation for electrons and for ions. It has been shown by a multiple
space time scale analysis that the dynamics is dominated by two
nonlinearities: the Charney-Hasegawa-Mima type, or vectorial nonlinearity,
generated by the ion polarization drift and the Korteweg-De Vries, or scalar
nonlinearity, related to the space variation of the density gradient length.
The former is of high differential degree and is dominant at small spatial
scales, of the order of few sonic Larmor radii $\rho _{s}$. The latter is
dominant at ``mesoscopic'' spatial scales, of the order of $\sqrt{\rho
_{s}L_{n}}$. The numerical studies \cite{Mikhailovskaya}, the fluid-tank
experiments \cite{Nezlin} and the multiple space-time scale analytical
analysis \cite{Spatschek3} show that the scalar nonlinearity becomes
dominant at late regimes in the statistical stationarity of the drift wave
turbulence. However the possible manifestation of the two types of
nonlinearity rises the problem of \emph{structural} stability of either
regime where only one of these nonlinearity is considered dominant:
inclusion of the other strongly changes the behavior of the system. It is
then pertinent to consider that the turbulence generated in a realistic
regime may include manifestations of both types. The turbulence is dominated
by the larger scales sustained by the scalar nonlinearity (described by the
Flierl-Petviashvili equation \cite{HH}) together with robust vortices
generated at the scales of few $\rho _{s}$, typical the CHM (\emph{i.e.}
vectorial) nonlinearity \cite{Nycander2}.

When the scalar nonlinearity is prevailing \cite{MH}, \cite{MH2} the
equation has the form

\begin{equation}
\left( 1-\nabla _{\perp }^{2}\right) \frac{\partial \varphi }{\partial t}%
+v_{\ast }\frac{\partial \varphi }{\partial y}-v_{\ast T}\varphi \frac{%
\partial \varphi }{\partial y}=0  \label{eqMH}
\end{equation}
In a moving frame and restricting to stationarity we obtain 
\begin{equation}
\nabla _{\perp }^{2}\varphi -\alpha \varphi -\beta \varphi ^{2}=0  \label{FP}
\end{equation}
The physical parameters are \cite{Lakhin}, \cite{LaedkeSpatschek1}, \cite{HH}
\begin{equation}
\alpha =\frac{1}{\rho _{s}^{2}}\left( 1-\frac{v_{d}}{u}\right) \;,\;\beta =%
\frac{c_{s}^{2}}{2u^{2}}\frac{\partial }{\partial x}\left( \frac{1}{L_{n}}%
\right)  \label{alphabeta}
\end{equation}
where $\rho _{s}=c_{s}/\Omega _{i}$, $c_{s}=\left( T_{e}/m_{i}\right) ^{1/2}$
and the potential is scaled as $\phi \rightarrow e\phi /T_{e}$. Here $L_{n}$
and $L_{T}$ are respectively the gradient lengths of the density and
temperature. The velocity is the diamagnetic velocity $v_{d}=\rho
_{s}c_{s}/L_{n}$. The condition for the validity of this equation are: $%
\left( k_{x}\rho _{s}\right) \left( k\rho _{s}\right) ^{2}\ll \eta _{e}\rho
_{s}/L_{n}$, where $\eta _{e}=L_{n}/L_{T_{e}}$. The coefficients $\alpha $
and $\beta $ have the dimension $\left( length\right) ^{-2}$. This form will
be used below.

\subsection{The structures}

The exact solution of the equation is 
\begin{eqnarray}
\varphi _{s}\left( y,t;y_{0},u\right) &=&-3\left( \frac{u}{v_{d}}-1\right) 
\notag \\
&&\hspace*{-1cm}\times \sec \mathrm{h}^{2}\left[ \frac{1}{2\rho _{s}}\left(
1-\frac{v_{d}}{u}\right) ^{1/2}\,\left( y-y_{0}-ut\right) \right]
\label{solMH}
\end{eqnarray}
where the velocity is restricted to the intervals $u>v_{d}$\ \ \ or\ \ \ $%
u<0 $. In Ref.\cite{MH} the radial extension of the solution is estimated
as: $\left( \Delta x\right) ^{2}\sim \rho _{s}L_{n}$. In our work we shall
assume that $u$ is close to $v_{d}$ , $u\gtrsim v_{d}$ (i.e. the structures
have small amplitudes). The monopolar vortex in this regime is discussed in
Ref. \cite{Nycander2}. For asymptotic form of the CHM equation see Ref.\cite
{FlorinMadi5}. We will adopt the one-dimensional section of the solution (%
\emph{i.e.} Eq.(\ref{solMH})) when we calculate the eigenmodes of the
determinant of the second functional derivative of the action. In
conclusion, we consider coherent structures which are monopolar vortices of
both signs of vorticity, with equal magnitudes and with random positions in
plane.

\section{Statistical analysis of the physical system}

\subsection{General functional framework}

The general method for constructing the action functional for a classical
stochastic system is described by Martin, Siggia and Rose (MSR) \cite{MSR},
and in path integral formalism, by Jensen \cite{Jensen}. Two reviews by
Krommes are very useful references on this point \cite{Krommes1}, \cite
{Krommes2}. The functional method has been applied in several concrete
problems and references may be consulted for details \cite{Flmadi1}, \cite
{Flmadi2}, \cite{FlorinMadi4}, \cite{FlorinMadi3}, \cite{FlorinMadi1}. We
here review few elementary procedures (see \cite{Amit}).

Consider a differential equation $F\left[ \phi \right] =0$ whose solution is 
$\phi ^{z}$. The unknown function belongs to a space of functions $\phi
\left( x,y\right) $. We want to select from this space of functions
precisely the one that is the solution of the differential equation and for
this we can use the functional Dirac $\delta $ : $\delta \left( \phi -\phi
^{z}\right) $. This can be represented as a product of usual $\delta $
functions in every point of space 
\begin{equation}
\delta \left( \phi -\phi ^{z}\right) =\prod_{k=1}^{N}\delta \left[ \phi
\left( x_{k}\right) -\phi ^{z}\left( x_{k}\right) \right]  \label{eq21}
\end{equation}
Any operation that will be done on a functional of $\phi $ can be now
particularized to the solution $\phi ^{z}$ by simply inserting this Dirac
functional. For example the calculation of a functional $\Omega \left( \phi
\right) $ at the function $\phi ^{z}$ can be done by a functional
integration over the space of all functions, with insertion of this $\delta $
functional. Using the Fourier representation of the ordinary Dirac functions
and going to the continuous limit we note the appearence of the dual
function $\chi $ 
\begin{eqnarray}
\Omega \left( \phi ^{z}\right) &=&\int \mathcal{D}\left[ \phi \right] \Omega
\left( \phi \right) \left( \left| \frac{\delta F}{\delta \phi }\right|
_{\phi ^{z}}\right) \delta \left[ F\left( \phi \right) \right]  \notag \\
&=&\left( \left| \frac{\delta F}{\delta \phi }\right| _{\phi ^{z}}\right)
\int \mathcal{D}\left[ \phi \right] \mathcal{D}\left[ \chi \right] \Omega
\left( \phi \right)  \notag \\
&&\times \exp \left\{ i\int dx\chi \left( x\right) F\left[ \phi \left(
x\right) \right] \right\}  \label{eq24}
\end{eqnarray}

We will define the partition function as usual, by the functional integral
of Boltzmann weights calculated on the base of the MSR action 
\begin{equation}
Z\left[ J\right] =\int \mathcal{D}\left[ \phi \right] \mathcal{D}\left[ \chi %
\right] \exp \left\{ \int dxdy\left( \chi F\left[ \phi \right] +J_{\phi
}\phi +J_{\chi }\chi \right) \right\}  \label{eq2408}
\end{equation}
The functional integration takes into account the fluctuations of the
physical field $\phi $ and of its dual $\chi $. The ``free-energy''
functional is defined by $\exp \left\{ W\left[ J\right] \right\} =Z\left[ J%
\right] $ from which the irreducible correlations (cumulants) are calculated
by functional derivatives to $J$. 
\begin{equation}
\left\langle \phi \left( x,y\right) \right\rangle =\left. \exp \left\{ -W 
\left[ J\right] \right\} \frac{\delta }{\delta J_{\phi }\left( x,y\right) }%
\exp \left\{ W\left[ J\right] \right\} \right| _{J=0}  \label{onec}
\end{equation}
and similar for higher cumulants. The two-point irreducible correlation for
the field is 
\begin{eqnarray}
&&\left\langle \phi \left( x,y\right) \phi \left( x^{\prime },y^{\prime
}\right) \right\rangle  \notag \\
&=&\int \mathcal{D}\left[ \phi \right] \mathcal{D}\left[ \chi \right] \phi
\left( x,y\right) \phi \left( x^{\prime },y^{\prime }\right)  \notag \\
&&\left. \times \exp \left\{ \int dxdy\left( \chi F\left[ \phi \right]
+J_{\phi }\phi +J_{\chi }\chi \right) \right\} \right| _{J=0}  \notag \\
&=&\frac{1}{Z\left[ J=0\right] }\left. \frac{\delta }{\delta J_{\phi }\left(
x,y\right) }\frac{\delta }{\delta J_{\phi }\left( x^{\prime },y^{\prime
}\right) }Z\left[ J\right] \right| _{J=0}  \label{twoc}
\end{eqnarray}

This is the general analytical instrument that will be used in the following
calculations. The calculations from this work are presented in greater
detail in Ref. \cite{FlorinMadi6}.

\section{Vortices with random positions}

\subsection{The discrete set of $N$ vortices in plane}

The action of the discrete set of vortices is determined by the sum of the
actions of the individual vortices plus a part that results from the
interaction between them \cite{Schaposnik}. The first part is simply the
time integration of the energy, \emph{i.e.} (since time factorizes) the
space integration of the expression of the product of the static potentials
associated with a single vortex, Eq.(\ref{solMH}), and to its dual, which in
the end simply means the square of the wave-form of the vortex potential.
This quantity is repeated for each of the $N$ vortices.

The partition function is 
\begin{eqnarray}
Z_{V} &=&\frac{1}{N!}\left( \prod_{j=1}^{N}Z_{V}^{\left( 0\right) }\right)
\sum_{\left\{ \alpha \right\} }\int \left( \prod_{a=1}^{N}\frac{1}{A}d%
\mathbf{R}^{\left( a\right) }\right)  \notag \\
&&\hspace*{-1cm}\hspace*{-0.5cm}\times \exp \left[ -\pi \underset{a>b}{%
\sum_{a=1}^{N}\sum_{b=1}^{N}}\int dxdy\int dx^{\prime }dy^{\prime }\rho
_{\omega }^{\left( a\right) }\left( x,y\right) \right.  \notag \\
&&\left. \times K_{0}\left( \rho _{s}^{-1}\left| \mathbf{R}^{\left( a\right)
}-\mathbf{R}^{\left( b\right) }\right| \right) \rho _{\omega }^{\left(
b\right) }\left( x^{\prime },y^{\prime }\right) \right]  \label{eq25}
\end{eqnarray}
with the following meaning.

The first factor simply takes into account $N$ independent vortices with
arbitrary positions in plane and expresses the fact that this part of the
partition function results from a Cartesian product of the $N$ statistical
ensembles, one for each vortex. The factor $N!$ takes into account the
permutation symmetry. The generating functional for a static vortex with
structure given by the interaction with random waves, is calculated in Eq.(%
\ref{cor5}) below. We have $Z_{V}^{\left( 0\right) }=Z_{V\varphi }\left[ J%
\right] $ where we have indicated by the index $V\varphi $ that the
partition function of the vortex includes the interaction vortex-turbulence,
and that the expression depends on the external current $J$, and will
contribute to any correlation that we will obtain by functional derivations
at $J$.

The sum is over the set of configurations $\left\{ \alpha \right\} $
characterized by random choices of positive and negative vortices.

The integrations over the positions of the centers $\mathbf{R}^{\left(
a\right) }$ of the vortices, $a=1,N$, express the fact that we allow
arbitrary positions in plane, with equal probability. Each integral is
normalized with the area of the physically interesting two-dimensional
region of the plane, $A$.

The exponent of the Boltzmann weight contains action resulting from the
interaction between vortices. We expect that for a dilute gas of vortices,
where the distance between the centers $\mathbf{R}^{\left( a\right) }-%
\mathbf{R}^{\left( b\right) }$ is much larger than the core diameter $d$, $%
\mathbf{R}^{\left( a\right) }-\mathbf{R}^{\left( b\right) }\gg d\;,\;a,b=1,N$
, the interaction is very weak. In order to describe the interaction between
vortices we start from the well-known alternative model of the drift waves
sustained by the ion polarization drift nonlinearity \cite{HH} \cite
{Morikawa}, \cite{Stewart}. In this model it is considered a set of $%
N_{\omega }$ point-like vortices of strength $\omega _{i}$ interacting in
plane by a short range potential expressed as the function $K_{0}$ (modified
Bessel function) of the relative distance between vortices. The potential $%
\phi ^{p}$ in a point $\mathbf{R}$ is a sum of contributions from all the $%
N_{\omega }$ point-vortices $\phi ^{p}\left( \mathbf{R}\right)
=\sum_{i=1}^{N_{\omega }}\omega _{i}K_{0}\left( \rho _{s}^{-1}\left| \mathbf{%
R-R}_{i}\right| \right) $ and the equations of motion $d\mathbf{R}_{i}/dt=-%
\mathbf{\nabla }\phi ^{p}\times \widehat{\mathbf{e}}_{z}$ (where $\widehat{%
\mathbf{e}}_{z}$ is the versor perpendicular on the plane). The distribution
of vorticity of the physical system (in particular the quasi-coherent
vortical structures) represents spatial variations of density of these
point-like vortices. The interaction between the \emph{physical} vortices
will result from the interaction between the point-like vortices, taking
into account the density of these objects. The energy of interaction is 
\begin{equation*}
H=\underset{i>j}{\sum_{i=1}^{N_{\omega }}\sum_{j=1}^{N_{\omega }}}\omega
_{i}\omega _{j}K_{0}\left( \rho _{s}^{-1}\left| \mathbf{R}_{i}-\mathbf{R}%
_{j}\right| \right)
\end{equation*}
called the Kirchhoff function. The range of spatial decay of the interaction
is the Larmor sonic radius $\rho _{s}$, which however may be modified to an 
\emph{effective} Larmor radius, in the presence of gradients and flow. When
we approach the continuum limit $N_{\omega }\rightarrow \infty $ the
envelope of the density becomes the physical vorticity $\omega \left(
x,y\right) $ which, for this stage of the problem is sufficient to be
considered as highly concentrated in the cores of the physical vortices and
almost vanishing in the rest. Taking the elementary point-vortices of equal
strength $\omega _{j}\equiv \omega _{0}$ we have that each physical vortex
is an integer multiple $N^{\left( a\right) }$ of this quantity. Now we will
associate with each physical vortex a continuous function, \emph{i.e.} its
vorticity defined on the whole plane, $\rho _{\omega }^{\left( a\right)
}\left( x,y\right) $, which is, as said, concentrated in $\left( x,y\right) $%
\begin{equation}
\rho _{\omega }^{\left( a\right) }\left( x,y\right) =N^{\left( a\right)
}\omega _{0}\delta \left( \mathbf{R}-\mathbf{R}^{\left( a\right) }\right)
\label{eq265}
\end{equation}
Then the energy is \cite{Schaposnik} 
\begin{eqnarray}
H &=&\underset{a>b}{\sum_{a=1}^{N}\sum_{b=1}^{N}}\int dxdy\int dx^{\prime
}dy^{\prime }  \notag \\
&&\times \rho _{\omega }^{\left( a\right) }\left( x,y\right) K_{0}\left( 
\frac{\left| \mathbf{R}^{\left( a\right) }-\mathbf{R}^{\left( b\right)
}\right| }{\rho _{s}}\right) \rho _{\omega }^{\left( b\right) }\left(
x^{\prime },y^{\prime }\right)  \label{eq27}
\end{eqnarray}
Due to Eq.(\ref{eq265}) the interaction energy is only the interaction
between the \emph{centers} $R^{\left( a\right) }$ and $R^{\left( b\right) }$
of the vortices. The summation proceeds by grouping the point-vortices into
physical vortices, then assuming that these (for only this stage of the
problem) have $\delta $-function shape and finally formally replacing this $%
\delta $ with a continuous distribution $\rho _{\omega }^{\left( a\right)
}\left( x,y\right) $. In this operation a number of infinities arise from
the energy of the interaction of the point-vortices which are grouped into
one physical vortex, since the relative distances are zero for them. This
singular part can be removed since it does not participate to the functional
variations induced by the ``external excitation'' current $J$.

\bigskip

For simplification of the computation we now only consider physical vortices
of equal amplitude (positive or negative) and then the vorticity
distribution $\rho _{\omega }^{\left( a\right) }\left( x,y\right) $ has
amplitude $\omega _{v}=p\omega _{0}$ ($p$ is an integer), multiplied by the
integer $n^{\left( a\right) }$ which can take the values $\pm 1$ for
positive or negative vorticity. We have $N^{\left( a\right) }=pn^{\left(
a\right) }$. The sum over the physical vortices' positions suggests to
define a formal unique function of vorticity $\rho _{\omega }\left(
x,y\right) $ for all the $N$ physical vortices 
\begin{equation*}
\rho _{\omega }\left( x,y\right) \equiv \sum_{a=1}^{N}\rho _{\omega
}^{\left( a\right) }\left( x,y\right) =\sum_{a=1}^{N}\omega _{v}n^{\left(
a\right) }\delta \left( \mathbf{R-R}^{\left( a\right) }\right) 
\end{equation*}
Further, the energy is normalized with a constant dimensional factor. The
interaction part can be rewritten 
\begin{eqnarray}
&&\hspace*{-0.9cm}\exp \left[ -\frac{\pi }{\rho _{s}^{4}\omega _{v}^{2}}%
\underset{a>b}{\sum_{a=1}^{N}\sum_{b=1}^{N}}\int dxdy\int dx^{\prime
}dy^{\prime }\right.   \notag \\
&&\left. \times \rho _{\omega }^{\left( a\right) }\left( x,y\right)
K_{0}\left( \rho _{s}^{-1}\left| \mathbf{R}^{\left( a\right) }-\mathbf{R}%
^{\left( b\right) }\right| \right) \rho _{\omega }^{\left( b\right) }\left(
x^{\prime },y^{\prime }\right) \right]   \notag \\
&=&\exp \left[ -\frac{1}{2\rho _{s}^{4}\omega _{v}^{2}}\int dxdy\int
dx^{\prime }dy^{\prime }\rho _{\omega }\left( x,y\right) G\left( \mathbf{R-R}%
^{\prime }\right) \rho _{\omega }\left( x^{\prime },y^{\prime }\right) %
\right]   \label{eq28}
\end{eqnarray}
where $G$ is the kernel of interaction, 
\begin{eqnarray}
G\left( \mathbf{R-R}^{\prime }\right)  &\equiv &\frac{1}{2\pi }\underset{a>b%
}{\sum \sum }K_{0}\left( \rho _{s}\left| \mathbf{R}^{\left( a\right) }-%
\mathbf{R}^{\left( b\right) }\right| \right)   \notag \\
&&\times \delta \left( \mathbf{R-R}^{\left( a\right) }\right) \delta \left( 
\mathbf{R}^{\prime }\mathbf{-R}^{\left( b\right) }\right)   \label{eq29}
\end{eqnarray}
The differential equation for $K_{0}$ is $\left( \Delta -1/\rho
_{s}^{2}\right) K_{0}\left( r/\rho _{s}\right) =-2\pi \delta \left( \mathbf{r%
}\right) $ . This helps to replace the interaction part with a Gaussian
functional integral, by introducing an auxiliary field $\psi $ 
\begin{eqnarray}
&&\exp \left[ -\frac{1}{2\rho _{s}^{4}\omega _{v}^{2}}\int dxdy\int
dx^{\prime }dy^{\prime }\rho _{\omega }\left( \mathbf{R}\right) G\left( 
\mathbf{R-R}^{\prime }\right) \rho _{\omega }\left( \mathbf{R}^{\prime
}\right) \right]   \notag \\
&=&p_{1}^{-1}\int \mathcal{D}\left[ \psi \right] \exp \left\{ -\frac{1}{2}%
\int dxdy\left[ \left( \mathbf{\nabla }\psi \right) ^{2}+\frac{1}{\rho
_{s}^{2}}\psi ^{2}\right] \right\}   \notag \\
&&\times \exp \left[ i\frac{1}{\rho _{s}^{2}\omega _{v}}\int dxdy\rho
_{\omega }\left( \mathbf{R}\right) \psi \left( \mathbf{R}\right) \right] 
\label{patrat}
\end{eqnarray}
with $p_{1}$ a normalization constant. We make a change of variable in the
functional integration $2\pi \psi \rightarrow \chi $ (this also changes the
normalization constant $p_{1}\rightarrow p$) and return to the partition
function Eq.(\ref{eq25}) 
\begin{eqnarray}
Z_{V} &=&p^{-1}\int \mathcal{D}\left[ \chi \right] \exp \left\{ -\frac{1}{%
8\pi ^{2}}\int dxdy\left[ \left( \mathbf{\nabla }\chi \right) ^{2}+\frac{1}{%
\rho _{s}^{2}}\chi ^{2}\right] \right\}   \notag \\
&&\times \frac{1}{N!}\left( Z_{V}^{\left( 0\right) }\right)
^{N}\sum_{\left\{ \alpha \right\} }\int \left( \prod_{a=1}^{N}\frac{1}{A}d%
\mathbf{R}^{\left( a\right) }\right) \exp \left[ i\sum_{a}n^{\left( a\right)
}\chi \left( \mathbf{R}^{\left( a\right) }\right) \right]   \label{eq32}
\end{eqnarray}
In the last factor we note that in the sum each term consists of two
contributions, corresponding to positive and negative vorticity, $n^{\left(
a\right) }=\pm 1$, and they are weighted with the same factor, $1/2$ 
\begin{eqnarray}
&&\frac{1}{N!}\left( Z_{V}^{\left( 0\right) }\right) ^{N}\sum_{\left\{
\alpha \right\} }\int \prod_{a=1}^{N}\frac{1}{A}d\mathbf{R}^{\left( a\right)
}\exp \left[ i\sum_{a}n^{\left( a\right) }\chi \left( \mathbf{R}^{\left(
a\right) }\right) \right]   \notag \\
&=&\frac{\left( Z_{V}^{\left( 0\right) }\right) ^{N}}{N!}\left[
\prod_{a=1}^{N}\int \frac{1}{A}d\mathbf{R}\frac{1}{2}\left( \exp \left[
i\chi \left( \mathbf{R}\right) \right] +\exp \left[ -i\chi \left( \mathbf{R}%
\right) \right] \right) \right]   \label{eq33}
\end{eqnarray}
In the last line we have removed the upper index $\left( a\right) $ since
all factors in the product are now identical. For a fixed number $N$ of
vortices the partition function $Z_{V}^{\left( 0\right) }$ (with
nonvanishing contribution to the derivatives to $J$) is decoupled from the
other factors and will provide $N$-times the same contribution. The other
factors, \emph{i.e.} the functional integral that contains the interaction
between the vortices can only appear in the final answer as a constant,
multiplying contributions coming from $Z_{V}^{\left( 0\right) }\left[ J%
\right] $. When $N$ is arbitrary the partition function must also include a
sum over terms each corresponding to a number $N$ of vortices 
\begin{eqnarray}
&&\sum_{N=0}^{\infty }\frac{\left( Z_{V}^{\left( 0\right) }\right) ^{N}}{N!}%
\left[ \int \frac{1}{A}d\mathbf{R}\frac{1}{2}\left( \exp \left[ i\chi \left( 
\mathbf{R}\right) \right] +\exp \left[ -i\chi \left( \mathbf{R}\right) %
\right] \right) \right] ^{N}  \notag \\
&=&\exp \left\{ \frac{Z_{V}^{\left( 0\right) }}{A}\int dxdy\cos \left[ \chi
\left( x,y\right) \right] \right\}   \label{eq35}
\end{eqnarray}
Then the partition function becomes 
\begin{equation}
Z_{V}=p^{-1}\int \mathcal{D}\left[ \chi \right] \exp \left\{ -\frac{1}{8\pi
^{2}}\int dxdy\left[ \left( \mathbf{\nabla }\chi \right) ^{2}+\frac{1}{\rho
_{s}^{2}}\chi ^{2}-\frac{8\pi ^{2}}{A}Z_{V}^{\left( 0\right) }\cos \left[
\chi \left( x,y\right) \right] \right] \right\}   \label{zcos}
\end{equation}
In this expression the quantity $Z_{V}^{\left( 0\right) }$ is a functional
integral over the space of fields $\phi _{s}\left( x,y\right) $ representing
a single vortex. In the absence of the background turbulence the field $\phi
_{s}\left( x,y\right) $ is a deterministic quantity [Eq.(\ref{solMH})] and
the statistical ensemble is trivially composed of one element. The
interaction with the background turbulence induce a fluctuating form and the
statistical properties can be obtained from $Z_{V}^{\left( 0\right) }$. In
other words, $Z_{V}$ includes two sources of \ fluctuations: one is the
fluctuation of the field of vorticity due to the random positions in plane
of the vortices (the gas of vortices) and the other is the fluctuation of
the shape of the generic vortex due to interaction with the turbulent
background.

If we want to use the expression (\ref{zcos}) as a generating functional for
correlation we must be able to drop into the functional integral the field
representing the vortices, \emph{i.e.} 
\begin{equation}
\phi _{V}\left( \mathbf{R}\right) =\sum_{a}\phi _{s}^{\left( a\right)
}\left( x,y\right)  \label{eq36}
\end{equation}
as in Eq.(\ref{twoc}). An external excitation by the current $J$ will
produce a change in $Z_{V}$ from the change of the vorticity of the gas of
vortices and from the change of $Z_{V}^{\left( 0\right) }$. We have 
\begin{equation}
\left\langle \phi _{V}\left( x,y\right) \right\rangle =\frac{1}{Z_{V}\left[
j=0\right] }\left[ \left( \frac{\delta Z_{V}}{\delta J\left( x,y\right) }%
\right) _{Vort}+\frac{\delta Z_{V}}{\delta Z_{V}^{\left( 0\right) }}\frac{%
\delta Z_{V}^{\left( 0\right) }}{\delta J\left( x,y\right) }\right]
\label{eq365}
\end{equation}
\begin{eqnarray}
\left\langle \phi _{V}\left( x,y\right) \phi _{V}\left( x^{\prime
},y^{\prime }\right) \right\rangle &=&\frac{1}{Z_{V}\left[ j=0\right] }\left[
\left( \frac{\delta ^{2}Z_{V}}{\delta J\left( x,y\right) \delta J\left(
x^{\prime },y^{\prime }\right) }\right) _{vort}\right.  \label{eq367} \\
&&+\frac{\delta ^{2}Z_{V}}{\delta \left( Z_{V}^{\left( 0\right) }\right) ^{2}%
}\frac{\delta Z_{V}^{\left( 0\right) }}{\delta J\left( x,y\right) }\frac{%
\delta Z_{V}^{\left( 0\right) }}{\delta J\left( x^{\prime },y^{\prime
}\right) }  \notag \\
&&\left. +\frac{\delta Z_{V}}{\delta Z_{V}^{\left( 0\right) }}\frac{\delta
^{2}Z_{V}^{\left( 0\right) }}{\delta J\left( x,y\right) \delta J\left(
x^{\prime },y^{\prime }\right) }\right]  \notag
\end{eqnarray}
The formulas are taken at $J=0$. The first terms in these equations are
related with the fluctuations of the vorticity of the gas of vortices as a
continuous version of the discrete set of physical vortices with arbitrary
positions in plane. The other terms are related to the fluctuation of the
shape of a vortex and in order to calculate these contribution we need the
explicit expression of $Z_{V}$, as a functional of $Z_{V}^{\left( 0\right) }$%
. Further, we will need the detailed expression of $Z_{V}^{\left( 0\right) }%
\left[ J\right] $ and this will be calculated in the next Section.

In order to obtain the contribution from the fluctuation of the gas of
vortices (the first terms in Eqs.(\ref{eq365}) and (\ref{eq367})), we
introduce a new term in the action, consisting of the interaction between
the vorticity distribution $\rho _{\omega }\left( \mathbf{R}\right) $ and an
external current, $J_{\omega }$%
\begin{equation}
i\frac{1}{\omega _{v}}\int dxdy\left[ \rho _{\omega }\left( x,y\right)
J_{\omega }\left( x,y\right) \right]  \label{eq670}
\end{equation}
(the factor $i$ is introduced for compatibility with Eq.(\ref{patrat})).
This current $J_{\omega }$ is an external excitation applied on the field of
the vorticity and not on the field of potential $\phi _{V}$ as we would need
in the Eq.(\ref{eq2044}). We may assume that there is a connection between
the current $J_{\omega }$ and the current $J$ (which acts on the field $\phi
_{V}$) but there is no need to specify this relation. Indeed, the Eq.(\ref
{eq2049}) shows that at the end both currents should be taken zero.

The last line of Eq.(\ref{patrat}) transforms as follows 
\begin{equation}
\exp \left[ i\frac{1}{\omega _{v}}\int dxdy\rho _{\omega }\left( \mathbf{R}%
\right) \psi \left( \mathbf{R}\right) \right] \rightarrow \exp \left\{ i%
\frac{1}{\omega _{v}}\int dxdy\rho _{\omega }\left( \mathbf{R}\right) \left[
\psi \left( \mathbf{R}\right) +J_{\omega }\left( \mathbf{R}\right) \right]
\right\}  \label{eq38}
\end{equation}
All the calculations following Eq.(\ref{patrat}) are repeated without
modification but in the last term, instead of the function $\chi \left(
x,y\right) $ we will have 
\begin{equation}
\cos \left[ \chi \left( x,y\right) \right] \rightarrow \cos \left[ \chi
\left( x,y\right) +J_{\omega }\left( x,y\right) \right]  \label{eq39}
\end{equation}
since this was the term which resulted from the presence of the function $%
\rho _{\omega }\left( \mathbf{R}\right) $ in the Eq.(\ref{patrat}). Making
the change of variable in the functional integration (of Jacobian $1$) 
\begin{equation}
\chi \rightarrow \chi -J_{\omega }  \label{eq40}
\end{equation}
the integrand in the action is expressed as 
\begin{equation}
-\frac{1}{8\pi ^{2}}\left\{ \left[ \mathbf{\nabla }\left( \chi -J_{\omega
}\right) \right] ^{2}+\frac{1}{\rho _{s}^{2}}\left( \chi -J_{\omega }\right)
^{2}-\frac{8\pi ^{2}}{A}Z_{V}^{\left( 0\right) }\cos \left[ \chi \left(
x,y\right) \right] \right\}  \label{eq41}
\end{equation}
The two-point correlation of the field of the vorticity fluctuations can be
calculated from 
\begin{equation}
\frac{1}{\omega _{v}^{2}}\left\langle \rho _{\omega }\left( x,y\right) \rho
_{\omega }\left( x^{\prime },y^{\prime }\right) \right\rangle =\left. \frac{1%
}{Z_{V}\left[ J_{\omega }=0\right] }\frac{\delta Z_{V}\left[ J\right] }{%
i\delta J_{\omega }\left( x,y\right) i\delta J_{\omega }\left( x^{\prime
},y^{\prime }\right) }\right| _{J_{\omega }=0}  \label{eq42}
\end{equation}
The squares in the action Eq.(\ref{eq41}) are expanded 
\begin{eqnarray}
- &&\frac{1}{8\pi ^{2}}\left\{ \left( \mathbf{\nabla }\chi \right) ^{2}+%
\frac{1}{\rho _{s}^{2}}\chi ^{2}-\frac{8\pi ^{2}}{A}Z_{V}^{\left( 0\right)
}\cos \left[ \chi \left( x,y\right) \right] \right.  \label{eq43} \\
&&+\left( \mathbf{\nabla }J_{\omega }\right) ^{2}+\frac{1}{\rho _{s}^{2}}%
J_{\omega }^{2}  \notag \\
&&\left. -2\left( \mathbf{\nabla }\chi \right) \left( \mathbf{\nabla }%
J_{\omega }\right) -\frac{2}{\rho _{s}^{2}}\chi J_{\omega }\right\}  \notag
\end{eqnarray}
In the part of the action that depends on $J_{\omega }$ we make integrations
by parts 
\begin{eqnarray}
&&\left( \mathbf{\nabla }J_{\omega }\right) ^{2}+\frac{1}{\rho _{s}^{2}}%
J_{\omega }^{2}-2\left( \mathbf{\nabla }\chi \right) \left( \mathbf{\nabla }%
J_{\omega }\right) -\frac{2}{\rho _{s}^{2}}\chi J_{\omega }  \label{eq44} \\
&\rightarrow &-J_{\omega }\left( \Delta J_{\omega }\right) +\frac{1}{\rho
_{s}^{2}}J_{\omega }^{2}+2J_{\omega }\left( \Delta \chi \right) -\frac{2}{%
\rho _{s}^{2}}\chi J_{\omega }  \notag
\end{eqnarray}
The first line at the exponent in Eq.(\ref{eq43}) does not contain the
current and in the following functional derivations to $J_{\omega }$ we will
temporary omit it. Consider the application of the first operator of
derivation to $iJ_{\omega }\left( x,y\right) $ 
\begin{eqnarray}
&&\hspace*{-0.9cm}\frac{\delta }{i\delta J_{\omega }\left( x,y\right) }\exp
\left\{ -\frac{1}{8\pi ^{2}}\int dxdy\left[ -J_{\omega }\left( \Delta
J_{\omega }\right) +\frac{1}{\rho _{s}^{2}}J_{\omega }^{2}\right. \right.
\label{eq45} \\
&&\left. \left. +2J_{\omega }\left( \Delta \chi \right) -\frac{2}{\rho
_{s}^{2}}\chi J_{\omega }\right] \right\}  \notag \\
&=&\frac{1}{i}\exp \left\{ ...\right\}  \notag \\
&&\times \rho _{s}^{2}\frac{\left( -1\right) }{8\pi ^{2}}\left( -2\Delta
J_{\omega }+\frac{2}{\rho _{s}^{2}}J_{\omega }+2\Delta \chi -\frac{2}{\rho
_{s}^{2}}\chi \right) _{\left( x,y\right) }  \notag
\end{eqnarray}
Every derivation to the current $J_{\omega }$ suppresses a space integration
and in consequence the result is multiplied with factors $\rho _{s}$ which
render the space integral dimensionless. The subscript shows that the
functions inside the bracket are calculated in the point $\left( x,y\right) $%
.

The second operator of derivation is now applied on Eq.(\ref{eq45}) 
\begin{eqnarray}
&&\frac{\delta }{i\delta J_{\omega }\left( x^{\prime },y^{\prime }\right) }%
\exp \left\{ -\frac{1}{8\pi ^{2}}\int dxdy\left[ -J_{\omega }\left( \Delta
J_{\omega }\right) +\frac{1}{\rho _{s}^{2}}J_{\omega }^{2}\right. \right.
\label{eq458} \\
&&\left. \left. +2J_{\omega }\left( \Delta \chi \right) -\frac{2}{\rho
_{s}^{2}}\chi J_{\omega }\right] \right\}  \notag \\
&&\times \frac{1}{i}\rho _{s}^{2}\frac{\left( -1\right) }{8\pi ^{2}}\left(
-2\Delta J_{\omega }+\frac{2}{\rho _{s}^{2}}J_{\omega }+2\Delta \chi -\frac{2%
}{\rho _{s}^{2}}\chi \right) _{\left( x,y\right) }  \notag \\
&=&\frac{1}{i}\rho _{s}^{2}\frac{\left( -1\right) }{8\pi ^{2}}\left(
-2\Delta J_{\omega }+\frac{2}{\rho _{s}^{2}}J_{\omega }+2\Delta \chi -\frac{2%
}{\rho _{s}^{2}}\chi \right) _{\left( x^{\prime },y^{\prime }\right) } 
\notag \\
&&\times \frac{1}{i}\rho _{s}^{2}\frac{\left( -1\right) }{8\pi ^{2}}\left(
-2\Delta J_{\omega }+\frac{2}{\rho _{s}^{2}}J_{\omega }+2\Delta \chi -\frac{2%
}{\rho _{s}^{2}}\chi \right) _{\left( x,y\right) }  \notag \\
&&\times \exp \left\{ ...\right\}  \notag \\
&&+\frac{1}{i}\frac{1}{i}\rho _{s}^{2}\frac{\left( -1\right) }{8\pi ^{2}}%
\left[ -2\Delta \delta \left( x-x^{\prime },y-y^{\prime }\right) +\frac{2}{%
\rho _{s}^{2}}\delta \left( x-x^{\prime },y-y^{\prime }\right) \right] 
\notag \\
&&\times \exp \left\{ ...\right\}  \notag
\end{eqnarray}
At this moment we can take $J\equiv 0$. The exponentials in the Eq(\ref
{eq458}) are equal to $1$. The auto-correlation is 
\begin{eqnarray}
&&\frac{1}{\omega _{v}^{2}}\left\langle \rho _{\omega }\left( x,y\right)
\rho _{\omega }\left( x^{\prime },y^{\prime }\right) \right\rangle
\label{eq459} \\
&=&\frac{1}{Z_{V}\left[ J_{\omega }=0\right] }p^{-1}\int \mathcal{D}\left[
\chi \right] \exp \left\{ -\frac{1}{8\pi ^{2}}\int dxdy\left[ \left( \mathbf{%
\nabla }\chi \right) ^{2}+\frac{1}{\rho _{s}^{2}}\chi ^{2}-\frac{8\pi ^{2}}{A%
}Z_{V}^{\left( 0\right) }\cos \left[ \chi \left( x,y\right) \right] \right]
\right\}  \notag \\
&&\times \left\{ \frac{\rho _{s}^{2}}{8\pi ^{2}}\left[ -2\Delta \delta
\left( x-x^{\prime },y-y^{\prime }\right) +\frac{2}{\rho _{s}^{2}}\delta
\left( x-x^{\prime },y-y^{\prime }\right) \right] \right.  \notag \\
&&\left. -\left( \frac{\rho _{s}^{2}}{8\pi ^{2}}\right) ^{2}\left( 2\Delta
\chi -\frac{2}{\rho _{s}^{2}}\chi \right) _{\left( x,y\right) }\left(
2\Delta \chi -\frac{2}{\rho _{s}^{2}}\chi \right) _{\left( x^{\prime
},y^{\prime }\right) }\right\}  \notag
\end{eqnarray}
The last two lines in Eq.(\ref{eq459}) (the curly bracket) arise from
derivation. The normalization gives by definition 
\begin{eqnarray*}
Z_{V}\left[ J_{\omega }=0\right] &\equiv &\left\langle 1\right\rangle \\
&=&p^{-1}\int \mathcal{D}\left[ \chi \right] \exp \left\{ -\frac{1}{8\pi ^{2}%
}\int dxdy\left[ \left( \mathbf{\nabla }\chi \right) ^{2}+\frac{1}{\rho
_{s}^{2}}\chi ^{2}-\frac{8\pi ^{2}}{A}Z_{V}^{\left( 0\right) }\cos \left[
\chi \left( x,y\right) \right] \right] \right\}
\end{eqnarray*}
The functional integration does not affect the first two terms in the curly
bracket in Eq.(\ref{eq459}). The functional integral of the last term
represents the average of the product of two functions $\chi $. 
\begin{eqnarray*}
&&\frac{1}{\omega _{v}^{2}}\left\langle \rho _{\omega }\left( x,y\right)
\rho _{\omega }\left( x^{\prime },y^{\prime }\right) \right\rangle \\
&=&\frac{\rho _{s}^{2}}{4\pi ^{2}}\left[ -\Delta \delta \left( x-x^{\prime
},y-y^{\prime }\right) +\frac{1}{\rho _{s}^{2}}\delta \left( x-x^{\prime
},y-y^{\prime }\right) \right] \\
&&-4\left( \frac{\rho _{s}^{2}}{8\pi ^{2}}\right) ^{2}\left( \Delta -\frac{1%
}{\rho _{s}^{2}}\right) _{\left( x,y\right) }\left( \Delta -\frac{1}{\rho
_{s}^{2}}\right) _{\left( x^{\prime },y^{\prime }\right) }\left\langle \chi
\left( x,y\right) \chi \left( x^{\prime },y^{\prime }\right) \right\rangle
\end{eqnarray*}

It is now easier if we use the Fourier representation of the fields, which
we denote by the symbol $\widetilde{}$. Writting $\mathbf{x}\equiv \left(
x,y\right) $, 
\begin{equation*}
\frac{1}{\omega _{v}^{2}}\left\langle \rho _{\omega }\left( x,y\right) \rho
_{\omega }\left( x^{\prime },y^{\prime }\right) \right\rangle =\frac{1}{%
\omega _{v}^{2}}\int d\mathbf{k}\exp \left( i\mathbf{k\cdot x}\right) \int d%
\mathbf{k}^{\prime }\exp \left( i\mathbf{k}^{\prime }\cdot \mathbf{x}%
^{\prime }\right) \left\langle \widetilde{\rho }_{\omega }\left( \mathbf{k}%
\right) \widetilde{\rho }_{\omega }\left( \mathbf{k}^{\prime }\right)
\right\rangle
\end{equation*}
We have 
\begin{eqnarray*}
&&\frac{1}{\omega _{v}^{2}}\int d\mathbf{k}\exp \left( i\mathbf{k\cdot x}%
\right) \int d\mathbf{k}^{\prime }\exp \left( i\mathbf{k}^{\prime }\cdot 
\mathbf{x}^{\prime }\right) \left\langle \widetilde{\rho }_{\omega }\left( 
\mathbf{k}\right) \widetilde{\rho }_{\omega }\left( \mathbf{k}^{\prime
}\right) \right\rangle \\
&=&\frac{\rho _{s}^{2}}{4\pi ^{2}}\int d\mathbf{k}\exp \left[ i\mathbf{%
k\cdot }\left( \mathbf{x-x}^{\prime }\right) \right] \left( k^{2}+\frac{1}{%
\rho _{s}^{2}}\right) \\
&&-\left( \frac{\rho _{s}^{2}}{4\pi ^{2}}\right) ^{2}\int d\mathbf{k}\exp
\left( i\mathbf{k\cdot x}\right) \int d\mathbf{k}^{\prime }\exp \left( i%
\mathbf{k}^{\prime }\cdot \mathbf{x}^{\prime }\right) \\
&&\times \left( k^{2}+\frac{1}{\rho _{s}^{2}}\right) \left( k^{\prime 2}+%
\frac{1}{\rho _{s}^{2}}\right) \left\langle \widetilde{\chi }\left( \mathbf{k%
}\right) \widetilde{\chi }\left( \mathbf{k}^{\prime }\right) \right\rangle
\end{eqnarray*}
We can take a fixed reference point 
\begin{eqnarray*}
\mathbf{x}^{\prime } &=&\mathbf{a} \\
\mathbf{x} &=&\mathbf{a+x-x}^{\prime }
\end{eqnarray*}
and write 
\begin{eqnarray}
&&\frac{1}{\omega _{v}^{2}}\int d\mathbf{k}\exp \left[ i\mathbf{k\cdot }%
\left( \mathbf{x-x}^{\prime }\right) \right] \int d\mathbf{k}^{\prime }\exp %
\left[ i\left( \mathbf{k}^{\prime }+\mathbf{k}\right) \cdot \mathbf{a}\right]
\left\langle \widetilde{\rho }_{\omega }\left( \mathbf{k}\right) \widetilde{%
\rho }_{\omega }\left( \mathbf{k}^{\prime }\right) \right\rangle
\label{eq4531} \\
&=&\frac{\rho _{s}^{2}}{4\pi ^{2}}\int d\mathbf{k}\exp \left[ i\mathbf{%
k\cdot }\left( \mathbf{x-x}^{\prime }\right) \right] \left( k^{2}+\frac{1}{%
\rho _{s}^{2}}\right)  \notag \\
&&-\left( \frac{\rho _{s}^{2}}{4\pi ^{2}}\right) ^{2}\int d\mathbf{k}\exp %
\left[ i\mathbf{k\cdot }\left( \mathbf{x-x}^{\prime }\right) \right] \int d%
\mathbf{k}^{\prime }\exp \left[ i\left( \mathbf{k}^{\prime }+\mathbf{k}%
\right) \cdot \mathbf{a}\right]  \notag \\
&&\times \left( k^{2}+\frac{1}{\rho _{s}^{2}}\right) \left( k^{\prime 2}+%
\frac{1}{\rho _{s}^{2}}\right) \left\langle \widetilde{\chi }\left( \mathbf{k%
}\right) \widetilde{\chi }\left( \mathbf{k}^{\prime }\right) \right\rangle 
\notag
\end{eqnarray}
The parameter $\mathbf{a}$ has no particular role : non of our assumption
has imposed a nonuniformity of the statistical properties on the plane.
Therefore we can integrate Eq.(\ref{eq4531}) over the position $\mathbf{a}$, 
\emph{i.e.} on the plane 
\begin{equation*}
\frac{1}{A}\int d\mathbf{a...}
\end{equation*}
Obviously, this will produce in the left hand side a function $\delta $%
\begin{equation*}
\delta \left( \mathbf{k}^{\prime }+\mathbf{k}\right)
\end{equation*}
after which the integration over the second wavenumber, $\mathbf{k}^{\prime
} $ , will impose 
\begin{equation*}
\mathbf{k}^{\prime }=-\mathbf{k}
\end{equation*}
For the first term in the right hand side, the integration over $\mathbf{a}$
will have no effect. For the second term the effect is the same as in the
left hand side, \emph{i.e.} we have $\mathbf{k}^{\prime }=-\mathbf{k}$. we
will now replace $\mathbf{x-x}^{\prime }$ by $\mathbf{x}$ and obtain 
\begin{eqnarray*}
&&\frac{1}{\omega _{v}^{2}}\int d\mathbf{k}\exp \left( i\mathbf{k\cdot x}%
\right) \left\langle \widetilde{\rho }_{\omega }\left( \mathbf{k}\right) 
\widetilde{\rho }_{\omega }\left( -\mathbf{k}\right) \right\rangle \\
&=&\frac{\rho _{s}^{2}}{4\pi ^{2}}\int d\mathbf{k}\exp \left( i\mathbf{%
k\cdot x}\right) \left( k^{2}+\frac{1}{\rho _{s}^{2}}\right) \\
&&-\left( \frac{\rho _{s}^{2}}{4\pi ^{2}}\right) ^{2}\int d\mathbf{k}\exp
\left( i\mathbf{k\cdot x}\right) \left( k^{2}+\frac{1}{\rho _{s}^{2}}\right)
^{2}\left\langle \widetilde{\chi }\left( \mathbf{k}\right) \widetilde{\chi }%
\left( -\mathbf{k}\right) \right\rangle
\end{eqnarray*}
In physical space the correlation also reflects the statistical uniformity, 
\begin{equation*}
\frac{1}{\omega _{v}^{2}}\int d\mathbf{k}\exp \left( i\mathbf{k\cdot x}%
\right) \left\langle \widetilde{\rho }_{\omega }\left( \mathbf{k}\right) 
\widetilde{\rho }_{\omega }\left( -\mathbf{k}\right) \right\rangle =\frac{1}{%
\omega _{v}^{2}}\left\langle \rho _{\omega }\left( \mathbf{x}\right) \rho
_{\omega }\left( \mathbf{0}\right) \right\rangle
\end{equation*}
The equation is 
\begin{eqnarray}
&&\frac{1}{\omega _{v}^{2}}\left\langle \widetilde{\rho }_{\omega }\left( 
\mathbf{k}\right) \widetilde{\rho }_{\omega }\left( -\mathbf{k}\right)
\right\rangle  \label{eq52} \\
&=&\frac{\rho _{s}^{2}}{4\pi ^{2}}\left( k^{2}+\frac{1}{\rho _{s}^{2}}%
\right) \left[ 1-\frac{\rho _{s}^{2}}{4\pi ^{2}}\left( k^{2}+\frac{1}{\rho
_{s}^{2}}\right) \left\langle \widetilde{\chi }\left( \mathbf{k}\right) 
\widetilde{\chi }\left( -\mathbf{k}\right) \right\rangle \right]  \notag
\end{eqnarray}
The second term in the bracket of Eq.(\ref{eq52}) contains the two-point
correlation of the function $\chi $ and can be obtained by explicit
calculation of the functional integration in Eq.(\ref{chichi}). The same
analytical problem as for the explicit calculation of $Z_{V}\left[ J=0\right]
$ and $\left\langle \chi \chi \right\rangle _{\mathbf{k}}\equiv \left\langle 
\widetilde{\chi }\left( \mathbf{k}\right) \widetilde{\chi }\left( -\mathbf{k}%
\right) \right\rangle $ will appear later (for the turbulence scattered by
the random vortices) and there we will give some details of calculation. At
this moment few explanations are sufficient. For small amplitude of the
auxiliary field $\chi $ the function $\cos $ is approximated with its first
two terms 
\begin{equation}
\cos \chi \approx 1-\frac{\chi ^{2}}{2}  \label{eq527}
\end{equation}
The constant $1$ is only a shift of the action. However it leads to a term
that depends on $Z_{V}^{\left( 0\right) }$, which is integrated in the
exponential over all volume \emph{i.e.} the area $A$ on the plane. In
detail, replacing Eq.(\ref{eq527}) in Eq.(\ref{eq525}) 
\begin{eqnarray}
Z_{V}\left[ J_{\omega }=0\right] &\equiv &p^{-1}\int \mathcal{D}\left[ \chi %
\right]  \label{eq528} \\
&&\times \exp \left[ -\frac{1}{8\pi ^{2}}\int dxdy\left( -\frac{8\pi ^{2}}{A}%
Z_{V}^{\left( 0\right) }\right) \right]  \notag \\
&&\times \exp \left\{ -\frac{1}{8\pi ^{2}}\int dxdy\left[ \left( \mathbf{%
\nabla }\chi \right) ^{2}+\frac{1}{\rho _{s}^{2}}\chi ^{2}+\left( \frac{8\pi
^{2}}{A}Z_{V}^{\left( 0\right) }\right) \frac{\chi ^{2}}{2}\right] \right\} 
\notag
\end{eqnarray}
The first factor can be taken outside the functional integration 
\begin{eqnarray}
&&\exp \left\{ -\frac{1}{8\pi ^{2}}\int dxdy\left( -\frac{8\pi ^{2}}{A}%
Z_{V}^{\left( 0\right) }\right) \right\}  \label{eq529} \\
&=&\exp \left[ Z_{V}^{\left( 0\right) }\right]  \notag
\end{eqnarray}
Since it is determined by the non-interacting vortices, it must exist even
if we would neglect completely the interaction between the vortices taking $%
\chi \rightarrow 0$. The rest of the Eq.(\ref{eq528}) is the Gaussian
functional integral 
\begin{eqnarray}
&&p^{-1}\int \mathcal{D}\left[ \chi \right]  \label{eq5221} \\
&&\times \exp \left\{ -\frac{1}{8\pi ^{2}}\int dxdy\left[ \left( \mathbf{%
\nabla }\chi \right) ^{2}+\frac{1}{\rho _{s}^{2}}\chi ^{2}+\left( \frac{8\pi
^{2}}{A}Z_{V}^{\left( 0\right) }\right) \frac{\chi ^{2}}{2}\right] \right\} 
\notag \\
&=&p^{-1}\int \mathcal{D}\left[ \chi \right] \\
&&\times \exp \left\{ -\frac{1}{8\pi ^{2}}\int dxdy\chi \left( x,y\right) %
\left[ -\Delta +\left( \frac{1}{\rho _{s}^{2}}+\frac{4\pi ^{2}}{A}%
Z_{V}^{\left( 0\right) }\right) \right] \chi \left( x,y\right) \right\} 
\notag \\
&=&q\left[ \det \left( -\Delta +\alpha ^{2}\right) \right] ^{-1/2}  \notag
\end{eqnarray}
where 
\begin{equation*}
\alpha ^{2}\equiv \frac{1}{\rho _{s}^{2}}+\frac{4\pi ^{2}}{A}Z_{V}^{\left(
0\right) }
\end{equation*}
The determinant can be calculated explicitely, by the product of the
eigenvalues of the operator. This product (besides an infinite factor that
will disappear) is convergent. However we keep this formal expression 
\begin{equation}
Z_{V}=Z_{V}\left[ J_{\omega }=0\right] =q\exp \left[ Z_{V}^{\left( 0\right) }%
\right] \left[ \det \left( -\Delta +\alpha ^{2}\right) \right] ^{-1/2}
\label{eq5227}
\end{equation}

The second term in the bracket of Eq.(\ref{eq52}) contains the two-point
correlation of the function $\chi $ and can be obtained by explicit
calculation of the functional integration in Eq.(\ref{eq459}). The same
analytical problem as for the explicit calculation of $Z_{V}\left[ J_{\omega
}=0\right] $ (Eq.(\ref{eq4592})) and $\left\langle \chi \chi \right\rangle _{%
\mathbf{k}}\equiv \left\langle \widetilde{\chi }\left( \mathbf{k}\right) 
\widetilde{\chi }\left( -\mathbf{k}\right) \right\rangle $ will appear later
(for the turbulence scattered by the random vortices) and there we will give
some details of calculation. At this moment few explanations are sufficient.
For small amplitude of the auxiliary field $\chi $ the function $\cos $ is
approximated with its first two terms 
\begin{equation}
\cos \chi \approx 1-\frac{\chi ^{2}}{2}
\end{equation}
The constant $1$ is only a shift of the action. However it leads to a term
that depends on $Z_{V}^{\left( 0\right) }$, which is integrated in the
exponential over all volume \emph{i.e.} the area $A$ on the plane. In
detail, replacing Eq.(\ref{eq527}) in Eq.(\ref{eq525}) 
\begin{eqnarray}
Z_{V}\left[ J_{\omega }=0\right] &\equiv &p^{-1}\int \mathcal{D}\left[ \chi %
\right] \exp \left[ -\frac{1}{8\pi ^{2}}\int dxdy\left( -\frac{8\pi ^{2}}{A}%
Z_{V}^{\left( 0\right) }\right) \right]  \notag \\
&&\times \exp \left\{ -\frac{1}{8\pi ^{2}}\int dxdy\left[ \left( \mathbf{%
\nabla }\chi \right) ^{2}+\frac{1}{\rho _{s}^{2}}\chi ^{2}+\left( \frac{8\pi
^{2}}{A}Z_{V}^{\left( 0\right) }\right) \frac{\chi ^{2}}{2}\right] \right\} 
\notag
\end{eqnarray}
The first factor can be taken outside the functional integration and is $%
\exp \left[ Z_{V}^{\left( 0\right) }\right] $. Since it is determined by the
non-interacting vortices, it must exist even if we would neglect completely
the interaction between the vortices taking $\chi \rightarrow 0$. The rest
of the Eq.(\ref{eq528}) is the Gaussian functional integral. Introducing the
notation $\alpha ^{2}\equiv 1/\rho _{s}^{2}+4\pi ^{2}Z_{V}^{\left( 0\right)
}/A$ 
\begin{equation}
Z_{V}\left[ J_{\omega }=0\right] =q\exp \left[ Z_{V}^{\left( 0\right) }%
\right] \left[ \det \left( -\Delta +\alpha ^{2}\right) \right] ^{-1/2}
\end{equation}
and $q$ is a constant. The determinant can be calculated explicitely, by the
product of the eigenvalues of the operator. We need this explicit expression
because we need the functional dependence of $Z_{V}=Z_{V}\left[ J_{\omega }=0%
\right] $ on $Z_{V}^{\left( 0\right) }$ as results from Eq.(\ref{eq367}). As
will become clear later, the factor with the determinant, which comes from
the influence of the fluctuating shape of a vortex on the correlations of
the vorticity of a gas of vortices in the plane, is affected by a factor $%
\rho _{s}^{2}/A$, which is small compared with the exponential in Eq.(\ref
{eq5227}). Therefore we calculate in a one dimensional cartezian
approximation the eigenvalues, instead of a cylindrical problem. We have to
solve 
\begin{equation*}
\left( -\frac{d^{2}}{dx^{2}}+\alpha ^{2}\right) \eta _{n}\left( x,y\right)
=\lambda _{n}^{\eta }\eta _{n}\left( x,y\right)
\end{equation*}
on an interval $L$. The eigenvalues are $\lambda _{n}^{\eta }=\left( 2\pi
n/L\right) ^{2}+\alpha ^{2}$ , where $n$ is an integer, and we obtain 
\begin{eqnarray}
\det \left( -\Delta +\alpha ^{2}\right) &=&\prod_{n=1}^{\infty }\lambda
_{n}^{\eta }=\prod_{n=1}^{\infty }\left[ \left( 2\pi n/L\right) ^{2}+\alpha
^{2}\right]  \label{eq5228} \\
&=&\prod_{n=1}^{\infty }\left( 2\pi n/L\right) ^{2}\prod_{n=1}^{\infty } 
\left[ 1+\frac{\alpha ^{2}L^{2}/\left( 2\pi \right) ^{2}}{n^{2}}\right] 
\notag \\
&=&\frac{\sinh \left( \alpha L/2\right) }{\alpha L/2}\prod_{n=1}^{\infty
}\left( 2\pi n/L\right) ^{2}  \notag
\end{eqnarray}
The infinite product is eliminated since we always use the ratios of $Z_{V}%
\left[ J_{\omega }\right] $ and $Z_{V}\left[ J_{\omega }=0\right] $. We also
take $L=\sqrt{A}$ and obtain 
\begin{equation}
Z_{V}=q\exp \left[ Z_{V}^{\left( 0\right) }\right] \left\{ \frac{\sinh \left[
\left( A\rho _{s}^{-2}+4\pi ^{2}Z_{V}^{\left( 0\right) }\right) ^{1/2}/2%
\right] }{\left( A\rho _{s}^{-2}+4\pi ^{2}Z_{V}^{\left( 0\right) }\right)
^{1/2}/2}\right\} ^{-1/2}  \label{eq5229}
\end{equation}
The quantity $A\rho _{s}^{-2}$ is very large and an approximation is
possible 
\begin{eqnarray}
&&\frac{\sinh \left[ \left( A\rho _{s}^{-2}+4\pi ^{2}Z_{V}^{\left( 0\right)
}\right) ^{1/2}/2\right] }{\left( A\rho _{s}^{-2}+4\pi ^{2}Z_{V}^{\left(
0\right) }\right) ^{1/2}/2}  \label{eq5230} \\
&=&\frac{1}{\rho _{s}^{-1}\sqrt{A}\left( 1+\frac{1}{2}\frac{4\pi
^{2}Z_{V}^{\left( 0\right) }}{A\rho _{s}^{-2}}\right) }\exp \left[ \frac{%
\rho _{s}^{-1}\sqrt{A}}{2}\left( 1+\frac{1}{2}\frac{4\pi ^{2}Z_{V}^{\left(
0\right) }}{A\rho _{s}^{-2}}\right) \right]  \notag \\
&\simeq &\frac{\exp \left( \rho _{s}^{-1}\sqrt{A}/2\right) }{\rho _{s}^{-1}%
\sqrt{A}}\exp \left( \frac{\pi ^{2}Z_{V}^{\left( 0\right) }}{\sqrt{A}\rho
_{s}^{-1}}\right) \left( 1-\frac{2\pi ^{2}Z_{V}^{\left( 0\right) }}{A\rho
_{s}^{-2}}\right)  \notag
\end{eqnarray}
The first factor is large but constant and can absorbed into the coefficient 
$q$. We get in Eq.(\ref{eq5229}) 
\begin{eqnarray*}
Z_{V} &=&q\exp \left[ Z_{V}^{\left( 0\right) }\right] \exp \left( -\frac{\pi
^{2}Z_{V}^{\left( 0\right) }}{2\sqrt{A}\rho _{s}^{-1}}\right) \left( 1-\frac{%
2\pi ^{2}Z_{V}^{\left( 0\right) }}{A\rho _{s}^{-2}}\right) ^{-1/2} \\
&=&q\exp \left[ Z_{V}^{\left( 0\right) }\left( 1-\frac{\pi ^{2}}{2\sqrt{A}%
\rho _{s}^{-1}}\right) \right] \left( 1+\frac{\pi ^{2}Z_{V}^{\left( 0\right)
}}{A\rho _{s}^{-2}}\right)
\end{eqnarray*}
We can neglect the second term in the first exponential 
\begin{equation}
Z_{V}=Z_{V}\left[ J_{\omega }=0\right] =q\left( 1+\frac{\pi
^{2}Z_{V}^{\left( 0\right) }}{A\rho _{s}^{-2}}\right) \exp \left[
Z_{V}^{\left( 0\right) }\right]  \label{eq5231}
\end{equation}

The second term in Eq.(\ref{eq52}), \emph{i.e.} the auto-correlation of $%
\chi $ in $\mathbf{k}$-space, may be calculated starting from the real-space
correlation 
\begin{eqnarray*}
&&\left\langle \chi \left( x,y\right) \chi \left( x^{\prime },y^{\prime
}\right) \right\rangle \\
&=&\frac{1}{Z_{V}\left[ J=0\right] }p^{-1}\exp \left[ Z_{V}^{\left( 0\right)
}\right] \int \mathcal{D}\left[ \chi \right] \chi \left( x,y\right) \chi
\left( x^{\prime },y^{\prime }\right) \\
&&\times \exp \left\{ -\frac{1}{8\pi ^{2}}\int dxdy\chi \left( x,y\right)
\left( -\Delta +\alpha ^{2}\right) \chi \left( x,y\right) \right\}
\end{eqnarray*}
As usual we return to the form of Eq.(\ref{eq5221}), and only for this step,
we insert an external current $J_{e}\left( x,y\right) $ interacting with $%
\chi $. The auxilliary functional is denoted $Z_{\chi }\left[ J_{e}\right] $ 
\begin{eqnarray*}
Z_{\chi }\left[ J_{e}\right] &=&\frac{1}{Z_{V}\left[ J=0\right] }p^{-1}\exp %
\left[ Z_{V}^{\left( 0\right) }\right] \int \mathcal{D}\left[ \chi \right] \\
&&\times \exp \left\{ -\frac{1}{8\pi ^{2}}\int dxdy\left[ \chi \left(
x,y\right) \left( -\Delta +\alpha ^{2}\right) \chi \left( x,y\right)
+J_{e}\chi \right] \right\} \\
&=&\frac{1}{Z_{V}\left[ J=0\right] }p^{-1}\exp \left[ Z_{V}^{\left( 0\right)
}\right] \int \mathcal{D}\left[ \phi \right] \\
&&\times \exp \left\{ -\frac{1}{8\pi ^{2}}\int dxdy\phi \left( x,y\right)
\left( -\Delta +\alpha ^{2}\right) \phi \left( x,y\right) \right. \\
&&\left. -\frac{1}{8\pi ^{2}}\int dxdy\frac{1}{4}J_{e}\left( x,y\right)
\left( -\Delta +\alpha ^{2}\right) ^{-1}J_{e}\left( x,y\right) \right\}
\end{eqnarray*}
To obtain the above equation we have made a change of variables $\chi
\rightarrow \phi =\chi +\frac{1}{2}\left( -\Delta +\alpha ^{2}\right)
^{-1}J_{e}$ of Jacobian $1$. The functional integration over $\phi $ can now
be carried out and the rest of the expression at the exponent appears in a
factor 
\begin{equation*}
\sim \left[ \det \left( -\Delta +\alpha ^{2}\right) \right] ^{-1/2}\exp
\left\{ -\frac{1}{8\pi ^{2}}\int dxdy\frac{1}{4}J_{e}\left( x,y\right)
\left( -\Delta +\alpha ^{2}\right) ^{-1}J_{e}\left( x,y\right) \right\}
\end{equation*}
where the symbol $\sim $ means that there also result constant factors. But
these are the same as those contained in the factor $q$ introduced in the
Eq.(\ref{eq5227}). We then have 
\begin{eqnarray*}
Z_{\chi }\left[ J_{e}\right] &=&\frac{1}{Z_{V}\left[ J=0\right] }p^{-1}\exp %
\left[ Z_{V}^{\left( 0\right) }\right] \int \mathcal{D}\left[ \chi \right] \\
&&\times \exp \left\{ -\frac{1}{8\pi ^{2}}\int dxdy\left[ \chi \left(
x,y\right) \left( -\Delta +\alpha ^{2}\right) \chi \left( x,y\right)
+J_{e}\chi \right] \right\} \\
&=&p^{-1}\exp \left\{ -\frac{1}{8\pi ^{2}}\int dxdy\frac{1}{4}J_{e}\left(
x,y\right) \left( -\Delta +\alpha ^{2}\right) ^{-1}J_{e}\left( x,y\right)
\right\}
\end{eqnarray*}
where we have taken into account Eq.(\ref{eq5227}). The correlation is 
\begin{eqnarray*}
&&\left\langle \chi \left( x,y\right) \chi \left( x^{\prime },y^{\prime
}\right) \right\rangle \\
&=&\frac{1}{Z_{\chi }\left[ J_{e}=0\right] }\left. \frac{\delta ^{2}Z_{\chi }%
\left[ J_{e}\right] }{\delta J_{e}\left( x,y\right) \delta J_{e}\left(
x^{\prime },y^{\prime }\right) }\right| _{J_{e}=0} \\
&=&\frac{1}{Z_{\chi }\left[ J_{e}=0\right] }p^{-1}\frac{\delta }{\delta
J_{e}\left( x^{\prime },y^{\prime }\right) }\left[ \left( -\frac{1}{8\pi ^{2}%
}2\left( -\Delta +\alpha ^{2}\right) ^{-1}J_{e}\left( x,y\right) \right)
\exp \left\{ ...\right\} \right] _{J_{e}=0} \\
&=&\frac{1}{Z_{\chi }\left[ J_{e}=0\right] }p^{-1}\left[ -\frac{1}{4\pi ^{2}}%
\left( -\Delta +\alpha ^{2}\right) ^{-1}\delta \left( \mathbf{x-x}^{\prime
}\right) \exp \left\{ ...\right\} \right. \\
&&\left. +\left( -\frac{1}{8\pi ^{2}}2\left( -\Delta +\alpha ^{2}\right)
^{-1}J_{e}\left( x,y\right) \right) \left( -\frac{1}{8\pi ^{2}}2\left(
-\Delta +\alpha ^{2}\right) ^{-1}J_{e}\left( x^{\prime },y^{\prime }\right)
\right) \exp \left\{ ...\right\} \right] _{J_{e}=0}
\end{eqnarray*}
(the factor $\exp \left[ Z_{V}^{\left( 0\right) }\right] $ has not been
written since it disappears). Finally we have 
\begin{equation}
\left\langle \chi \left( x,y\right) \chi \left( x^{\prime },y^{\prime
}\right) \right\rangle =-\frac{1}{4\pi ^{2}}\left( -\Delta +\alpha
^{2}\right) ^{-1}\delta \left( \mathbf{x-x}^{\prime }\right)  \label{eq53}
\end{equation}
or 
\begin{equation}
\left\langle \chi \chi \right\rangle _{\mathbf{k}}=\left[ \frac{1}{4\pi ^{2}}%
\rho _{s}^{2}\left( k^{2}+\frac{1}{\rho _{s}^{2}}+\frac{4\pi ^{2}}{A}%
Z_{V}^{\left( 0\right) }\right) \right] ^{-1}  \label{eq536}
\end{equation}
Introducing this in Eq.(\ref{eq52}) the correlation of the field of the
scalar potential will have from this a contribution (with normalization) 
\begin{eqnarray}
\frac{1}{\phi _{0}^{2}}\left\langle \phi _{V}\phi _{V}\right\rangle _{%
\mathbf{k}}^{vort} &=&\frac{1}{k^{2}\rho _{s}^{2}}\left( 1+\frac{1}{%
k^{2}\rho _{s}^{2}}\right)  \label{q54} \\
&&\hspace*{-0.9cm}\times \frac{1}{4\pi ^{2}}\left[ 1-\frac{1+k^{2}\rho
_{s}^{2}}{1+k^{2}\rho _{s}^{2}+Z_{V}^{\left( 0\right) }4\pi ^{2}\rho
_{s}^{2}/A}\right]  \notag
\end{eqnarray}
which represents, as said, the first term of the Eq.(\ref{eq367}).

The other terms in the correlation $\left\langle \phi _{V}\phi
_{V}\right\rangle $ (Eq.(\ref{eq367})) are due to the fluctuation of the
form of the generic vortex from interaction with the background turbulence.
We calculate, using Eq.(\ref{eq5237}), the derivatives 
\begin{equation}
\frac{\delta Z_{V}}{\delta Z_{V}^{\left( 0\right) }}=q\left( 1+\frac{\pi ^{2}%
}{A\rho _{s}^{-2}}+\frac{\pi ^{2}Z_{V}^{\left( 0\right) }}{A\rho _{s}^{-2}}%
\right) \exp \left[ Z_{V}^{\left( 0\right) }\right]  \label{eq5411}
\end{equation}
and 
\begin{equation}
\frac{\delta ^{2}Z_{V}}{\delta \left[ Z_{V}^{\left( 0\right) }\right] ^{2}}%
=q\left( 1+\frac{2\pi ^{2}}{A\rho _{s}^{-2}}+\frac{\pi ^{2}Z_{V}^{\left(
0\right) }}{A\rho _{s}^{-2}}\right) \exp \left[ Z_{V}^{\left( 0\right) }%
\right]  \label{eq5412}
\end{equation}

Now we have to calculate the derivatives of $Z_{V}^{\left( 0\right) }$ to $J$
as shown by the last two terms Eq.(\ref{eq367}). This part will be added
after $Z_{V}^{\left( 0\right) }$ is calculated, in the next subsection.

\subsection{A single vortex interacting with a turbulent environment}

The partition function for a single vortex in interaction with turbulence is
defined as 
\begin{equation}
Z_{V}^{\left( 0\right) }=\mathcal{N}^{-1}\int D\left[ \chi \right] D\left[
\phi \right] \exp \left\{ S_{V}\left[ \chi ,\phi \right] \right\}
\label{eq55}
\end{equation}
and the equation is the stationary form of the equation used in Ref.\cite
{FlorinMadi1}, 
\begin{equation}
F\left[ \phi \right] \equiv \nabla _{\perp }^{2}\varphi -\alpha \varphi
-\beta \varphi ^{2}  \label{eq57}
\end{equation}
The density of Lagrangean 
\begin{equation}
\mathcal{L}\left[ \chi \left( x,y\right) ,\phi \left( x,y\right) \right]
=\chi \left( \nabla _{\perp }^{2}\varphi -\alpha \varphi -\beta \varphi
^{2}\right)  \label{lv}
\end{equation}
is obtained from the Martin-Siggia-Rose method for classical stochastic
systems. Then the action is 
\begin{equation}
S_{V}\left[ \chi ,\phi \right] =\int dxdy\mathcal{L}\left[ \chi \left(
x,y\right) ,\phi \left( x,y\right) \right]  \label{sv}
\end{equation}
As usual we introduce the interaction with the external current $\mathbf{J=}%
\left( J_{\phi },J_{\chi }\right) $. However since there is no need to
calculate functional derivatives to $\chi $, we can only keep $J\equiv
J_{\phi }$. For uniformity of notation in this paper, we will not use the
factor $i$ in front of the action in contrast with \cite{FlorinMadi1}. We
have then 
\begin{equation}
Z_{V}^{\left( 0\right) }\left[ J\right] =\mathcal{N}^{-1}\int D\left[ \chi %
\right] D\left[ \phi \right] \exp \left\{ \int dxdy\left[ \chi \left( \nabla
_{\perp }^{2}\varphi -\alpha \varphi -\beta \varphi ^{2}\right) +J\phi %
\right] \right\}  \label{eq572}
\end{equation}

We need the explicit expression of the functional integral Eq.(\ref{eq572})
and this has been obtained in the references \cite{FlorinMadi1} and \cite
{FlorinMadi2}. For convenience we will recall briefly the steps of the
calculation restricting to the results we need in the present work.

To calculate the generating functional of the vortex in the background
turbulence we proceed in two steps: we solve the Euler-Lagrange equation for
the action (\ref{sv}) obtaining the configuration of the system which
extremises this action; further, we expand the action to second order in the
fluctuations around this extremum (which will include the turbulent field)
and integrate. The Euler-Lagrange equations have the solutions 
\begin{eqnarray}
\varphi _{Js}(x,y) &\equiv &\varphi _{s}\left( x,y\right)  \label{eq576} \\
\chi _{Js}\left( x,y\right) &=&-\,\varphi _{s}\left( x,y\right) +\widetilde{%
\chi }_{J}\left( x,y\right)  \notag
\end{eqnarray}
The first is the static form of the solution Eq.(\ref{solMH}) and does not
depend on $J$. The dual function is $-\,\varphi _{s}\left( x,y\right) $ plus
a term resulting from the excitation by $J$ in its equation. This additional
term $\widetilde{\chi }_{J}\left( x,y\right) $ is calculated by the
perturbation of the KdV soliton solution according to the modification of
the Inverse Scattering Transform when an inhomogeneous term (\emph{i.e.} $J$%
) is included. The action functional is calculated for these two functions 
\begin{equation*}
S_{Vs}\left[ J\right] \equiv S_{V}\left[ \varphi _{s}\left( x,y\right)
,-\,\varphi _{s}\left( x,y\right) +\widetilde{\chi }_{J}\left( x,y\right) %
\right]
\end{equation*}
Then the first part of our calculation is $Z_{V}^{\left( 0\right) }\left[ J%
\right] \sim \mathcal{N}^{-1}\exp \left\{ S_{Vs}\left[ J\right] \right\} $.
Expanding the action around this extremum 
\begin{equation*}
S_{V}\left[ \chi ,\varphi ;J\right] =S_{V}\left[ \varphi _{Js},\chi _{Js}%
\right] +\frac{1}{2}\left( \left. \frac{\delta ^{2}S_{V}\left[ J\right] }{%
\delta \varphi \delta \chi }\right| _{\varphi _{Js},\chi _{Js}}\right)
\delta \varphi \delta \chi
\end{equation*}
we calculate the Gaussian integral and obtain 
\begin{equation*}
Z_{V}^{\left( 0\right) }\left[ J\right] =\mathcal{N}^{-1}\exp \left\{ S_{Vs}%
\left[ J\right] \right\} \left[ \det \left( \left. \frac{\delta ^{2}S_{V}%
\left[ J\right] }{\delta \varphi \delta \chi }\right| _{\varphi _{Js},\chi
_{Js}}\right) \right] ^{-1/2}
\end{equation*}

If we can neglect the advection of vortices by large scale wave-like
fluctuations, we can calculate the determinant since the product of the
eigenvalues converges without the need for regularization. The result is 
\begin{equation}
Z_{V}^{\left( 0\right) }\left[ J\right] =\mathcal{N}^{-1}\exp \left\{ S_{Vs}%
\left[ J\right] \right\} \,A\,B  \label{cor5}
\end{equation}
Where 
\begin{equation}
A=A\left[ J\right] \equiv \left[ \frac{\beta /2}{\sinh \left( \beta
/2\right) }\right] ^{1/4}  \label{cor3}
\end{equation}
\begin{equation}
B=B\left[ J\right] \equiv \left[ \frac{\sigma /2}{\sin \left( \sigma
/2\right) }\right] ^{1/2}  \label{cor4}
\end{equation}
The eigenvalue problem depends functionally on $\widetilde{\chi }_{J}\left(
x,y\right) $ which implies that $\beta $ and $\sigma $ depend on $J$. Their
expressions can be found in \cite{FlorinMadi2}. With those detailed formulas
we can consider that we have the necessary knowledge to proceed to the
calculation of the functional derivatives of $Z_{V}^{\left( 0\right) }\left[
J\right] $ at $J$, using Eq.(\ref{cor5}). 
\begin{equation}
\frac{1}{Z_{V}^{\left( 0\right) }\left[ J=0\right] }\frac{\delta
Z_{V}^{\left( 0\right) }\left[ J\right] }{\delta J}=\frac{\delta S_{Vs}\left[
J\right] }{\delta J}+\frac{1}{A}\frac{\delta A}{\delta J}+\frac{1}{B}\frac{%
\delta B}{\delta J}  \label{eq585}
\end{equation}
For the present problem it is sufficient to take the first term as Eq.(\ref
{solMH}) 
\begin{equation}
\left. \frac{\delta S_{Vs}\left[ J\right] }{\delta J}\right| _{J=0}\simeq
\phi _{s}  \label{eq5852}
\end{equation}
The next two terms in Eq.(\ref{eq585}) represent the averaged, systematic
modification of the shape of the field around the vortex due to the mutual
interaction. They will be equally neglected, assuming that the main effect
is contained in the dispersion of the fluctuations of the shape of the
vortex interacting with the random field (which actually is our main concern
here). Then 
\begin{equation}
\frac{1}{Z_{V}^{\left( 0\right) }\left[ J=0\right] }\frac{\delta
Z_{V}^{\left( 0\right) }\left[ J\right] }{\delta J}\simeq \phi _{s}
\label{eq586}
\end{equation}

The second derivative to the excitations in two points $y_{1}$ and $y_{2}$
is 
\begin{eqnarray}
&&\text{ }\frac{1}{Z_{V}^{\left( 0\right) }\left[ J=0\right] }\frac{\delta
^{2}Z_{V}^{\left( 0\right) }\left[ J\right] }{\delta J\left( y_{2}\right)
\delta J\left( y_{1}\right) }  \label{eq587} \\
&=&\frac{\delta S_{Vs}\left[ J\right] }{\delta J\left( y_{2}\right) }\frac{%
\delta S_{Vs}\left[ J\right] }{\delta J\left( y_{1}\right) }+\frac{\delta
^{2}S_{Vs}\left[ J\right] }{\delta J\left( y_{2}\right) \delta J\left(
y_{1}\right) }  \notag \\
&&+\frac{1}{A}\frac{\delta A}{\delta J\left( y_{2}\right) }\frac{\delta
S_{Vs}\left[ J\right] }{\delta J\left( y_{1}\right) }+\frac{1}{B}\frac{%
\delta B}{\delta J\left( y_{2}\right) }\frac{\delta S_{Vs}\left[ J\right] }{%
\delta J\left( y_{1}\right) }  \notag \\
&&+\frac{1}{A}\frac{\delta A}{\delta J\left( y_{1}\right) }\frac{\delta
S_{Vs}\left[ J\right] }{\delta J\left( y_{2}\right) }+\frac{1}{B}\frac{%
\delta B}{\delta J\left( y_{1}\right) }\frac{\delta S_{Vs}\left[ J\right] }{%
\delta J\left( y_{2}\right) }  \notag \\
&&+\frac{1}{A}\frac{\delta A}{\delta J\left( y_{1}\right) }\frac{1}{B}\frac{%
\delta B}{\delta J\left( y_{2}\right) }+\frac{1}{A}\frac{\delta A}{\delta
J\left( y_{2}\right) }\frac{1}{B}\frac{\delta B}{\delta J\left( y_{1}\right) 
}  \notag \\
&&+\frac{1}{A}\frac{\delta ^{2}A}{\delta J\left( y_{2}\right) \delta J\left(
y_{1}\right) }+\frac{1}{B}\frac{\delta ^{2}B}{\delta J\left( y_{2}\right)
\delta J\left( y_{1}\right) }  \notag
\end{eqnarray}
Since we have assumed as an acceptable approximation to neglect the averaged
change produced by the turbulence on the soliton shape the second term in
the RHS is zero. For the first term we use Eq.(\ref{eq5852}). We have 
\begin{eqnarray*}
&&\frac{1}{Z_{V}^{\left( 0\right) }\left[ J=0\right] }\frac{\delta
^{2}Z_{V}^{\left( 0\right) }\left[ J\right] }{\delta J\left( y_{2}\right)
\delta J\left( y_{1}\right) } \\
&=&\phi _{s}\left( y_{2}\right) \phi _{s}\left( y_{1}\right) + \\
&&+\phi _{s}\left( y_{1}\right) \left[ \frac{\delta }{\delta J\left(
y_{2}\right) }\ln A+\frac{\delta }{\delta J\left( y_{2}\right) }\ln B\right]
\\
&&+\phi _{s}\left( y_{2}\right) \left[ \frac{\delta }{\delta J\left(
y_{1}\right) }\ln A+\frac{\delta }{\delta J\left( y_{1}\right) }\ln B\right]
\\
&&+\frac{\delta \ln A}{\delta J\left( y_{1}\right) }\frac{\delta \ln B}{%
\delta J\left( y_{2}\right) }+\frac{\delta \ln A}{\delta J\left(
y_{2}\right) }\frac{\delta \ln B}{\delta J\left( y_{1}\right) } \\
&&+\frac{1}{A}\frac{\delta ^{2}A}{\delta J\left( y_{2}\right) \delta J\left(
y_{1}\right) }+\frac{1}{B}\frac{\delta ^{2}B}{\delta J\left( y_{2}\right)
\delta J\left( y_{1}\right) }
\end{eqnarray*}
The expressions are complicated (see the Appendix of Ref.\cite{FlorinMadi2})
and some numerical calculation of these expression is unavoidable. For small
amplitude of the turbulent field the expression can be rewritten 
\begin{eqnarray*}
&&\left. \frac{1}{Z_{V}^{\left( 0\right) }\left[ J=0\right] }\frac{\delta
^{2}Z_{V}^{\left( 0\right) }\left[ J\right] }{\delta J\left( y_{2}\right)
\delta J\left( y_{1}\right) }\right| _{J=0} \\
&=&\phi _{s}\left( y_{2}\right) \phi _{s}\left( y_{1}\right) \left[
1+f\left( y\right) \right]
\end{eqnarray*}
\emph{i.e.} in a form that expresses the fact that the two-point correlation
is basically the auto-correlation of the potential of the exact soliton
modified by a function $f$ which collects the contributions from the
interaction with the random field. In $k$-space we have 
\begin{equation*}
\left. \frac{1}{Z_{V}^{\left( 0\right) }\left[ J=0\right] }\frac{\delta
^{2}Z_{V}^{\left( 0\right) }\left[ J\right] }{\delta J\left( y_{2}\right)
\delta J\left( y_{1}\right) }\right| _{J=0,\mathbf{k}}=\phi _{s}\left( 
\mathbf{k}\right) \phi _{s}\left( -\mathbf{k}\right) \left[ 1+f\left( 
\mathbf{k}\right) \right]
\end{equation*}

At the limit where we do not expand the action to include configurations
resulting from the interaction vortex-turbulence, we have $\sigma
\rightarrow 0$ and $\beta \rightarrow 0$ and it results $A=B=1$. In this
case $f\equiv 0$.

The detailed expressions of these terms are given in the paper \cite
{FlorinMadi1}, \cite{FlorinMadi2}. For the purpose of comparisons we will
express the spectrum as 
\begin{equation}
\frac{1}{\phi _{0}^{2}}\left\langle \phi _{V}\phi _{V}\right\rangle _{%
\mathbf{k}}=S\left( \mathbf{k}\right) \left[ 1+f\left( \mathbf{k}\right) %
\right]  \label{eq60}
\end{equation}
where $\phi _{0}$ is amplitude of a vortex, $S\left( \mathbf{k}\right) $ has
been derived by Meiss and Horton \cite{MH} 
\begin{equation}
S\left( \mathbf{k}\right) =\left\{ 12\sqrt{2}\pi ^{3/2}k\rho _{s}\frac{u}{%
v_{\ast }}\csc h\left[ \frac{\pi k\rho _{s}}{\left( 1-v_{\ast }/u\right)
^{1/2}}\right] \right\} ^{2}  \label{eq61}
\end{equation}
and $f\left( \mathbf{k}\right) $ is function that is the correction to the
Fourier transform of the squared secant-hyperbolic, produced by the
turbulent waves.

Before proceeding further with the calculations based on the Eq.(\ref{eq60})
we need to discuss the formal term $f\left( \mathbf{k}\right) $. Since this
term represents the difference from the simple isolated vortex to the vortex
perturbed by turbulence, one would like to have a quantitative connection
between the amplitude of this term and at least two elements characterizing
the background turbulence: (1) the amplitude and (2) the spectrum.

In the way we have conducted the calculations of the generating functional
Eqs.(\ref{cor5}), (\ref{cor3}), (\ref{cor4}) the new terms in the expression
of the auto-correlation due to the factors $A$ and $B$ are expressed in real
space, not in Fourier space. They are obtained from the product of
eigenvalues of the operator representing the second order functional
derivative of the action, \emph{i.e.} they are connected with the geometry
of the function space around the exact, vortex, nonlinear solution. The
determinant of the operator $\delta ^{2}S_{V}/\left( \delta \varphi \delta
\chi \right) $ may be seen as a volume in the function space, centered on
the vortex solution. The inverse of any eigenvalue gives an idea of the
extension along a particular direction (eigenfunction) in function space.
When an eigenvalue is very small, the operator almost vanishes on functions
along that direction. At the limit this is a \emph{zero mode} and
corresponds to a translational symmetry of the physical system along that
direction. The correlations depend on the sensitivity of this volume (the
product of the eigenvalues) on the excitation $J$ applied on the system. The
excitation is first manifested in the appearence of $\widetilde{\chi }%
_{J}\left( x,y\right) $. This one consists of a part that will modify the
shape of the exact vortex plus the oscillating tail generated when a soliton
is perturbed. The latter can be considered as a component of the background
turbulence. The ``propagation'' of the influence from an excitation $J$ can
be summarised symbolically in the chain : $J\rightarrow \widetilde{\chi }%
_{J}\left( x,y\right) \rightarrow $ eigenvalues of the operator $\delta
^{2}S_{V}/\left( \delta \varphi \delta \chi \right) \rightarrow $ $A$ and $B$
(or $\sigma $ and $\beta $). The expressions of $\sigma $ and $\beta $ can
be found in \cite{FlorinMadi2}. 

\bigskip

Now we can return to the Eq.(\ref{eq367}). Using Eq.(\ref{eq586}) the second
term is 
\begin{eqnarray}
&&\frac{1}{Z_{V}\left[ J=0\right] }\frac{\delta ^{2}Z_{V}}{\delta \left(
Z_{V}^{\left( 0\right) }\right) ^{2}}\frac{\delta Z_{V}^{\left( 0\right) }}{%
\delta J\left( x,y\right) }\frac{\delta Z_{V}^{\left( 0\right) }}{\delta
J\left( x^{\prime },y^{\prime }\right) }  \label{eq62} \\
&=&\frac{1}{\exp \left[ Z_{V}^{\left( 0\right) }\right] q\left(
1+Z_{V}^{\left( 0\right) }\rho _{s}^{2}\pi ^{2}/A\right) }\exp \left[
Z_{V}^{\left( 0\right) }\right] q\left[ 1+\left( 2+Z_{V}^{\left( 0\right)
}\right) \rho _{s}^{2}\pi ^{2}/A\right]  \notag \\
&&\times \phi _{s}\left( x,y\right) \phi _{s}\left( x^{\prime },y^{\prime
}\right)  \notag \\
&=&\left( 1+\frac{2\rho _{s}^{2}\pi ^{2}/A}{1+Z_{V}^{\left( 0\right) }\rho
_{s}^{2}\pi ^{2}/A}\right) \phi _{s}\left( x,y\right) \phi _{s}\left(
x^{\prime },y^{\prime }\right)  \notag
\end{eqnarray}
The third term in Eq.(\ref{eq367}) is 
\begin{eqnarray*}
&&\frac{1}{Z_{V}\left[ j=0\right] }\frac{\delta Z_{V}}{\delta Z_{V}^{\left(
0\right) }}\frac{\delta ^{2}Z_{V}^{\left( 0\right) }}{\delta J\left(
x,y\right) \delta J\left( x^{\prime },y^{\prime }\right) } \\
&=&\frac{1}{\exp \left[ Z_{V}^{\left( 0\right) }\right] q\left(
1+Z_{V}^{\left( 0\right) }\rho _{s}^{2}\pi ^{2}/A\right) }q\left( 1+\frac{%
\pi ^{2}}{A\rho _{s}^{-2}}+\frac{\pi ^{2}Z_{V}^{\left( 0\right) }}{A\rho
_{s}^{-2}}\right) \exp \left[ Z_{V}^{\left( 0\right) }\right] \\
&&\times \phi _{s}\left( x,y\right) \phi _{s}\left( x^{\prime },y^{\prime
}\right) \left( 1+f\right) \\
&=&\left( 1+\frac{\rho _{s}^{2}\pi ^{2}/A}{1+Z_{V}^{\left( 0\right) }\rho
_{s}^{2}\pi ^{2}/A}\right) \phi _{s}\left( x,y\right) \phi _{s}\left(
x^{\prime },y^{\prime }\right) \left( 1+f\right)
\end{eqnarray*}
In $\mathbf{k}$ space the two contributions reads 
\begin{eqnarray*}
&&\frac{1}{Z_{V}\left[ J=0\right] }\frac{\delta ^{2}Z_{V}}{\delta \left(
Z_{V}^{\left( 0\right) }\right) ^{2}}\frac{\delta Z_{V}^{\left( 0\right) }}{%
\delta J\left( x,y\right) }\frac{\delta Z_{V}^{\left( 0\right) }}{\delta
J\left( x^{\prime },y^{\prime }\right) }+\frac{1}{Z_{V}\left[ j=0\right] }%
\frac{\delta Z_{V}}{\delta Z_{V}^{\left( 0\right) }}\frac{\delta
^{2}Z_{V}^{\left( 0\right) }}{\delta J\left( x,y\right) \delta J\left(
x^{\prime },y^{\prime }\right) } \\
&=&\phi _{0}^{2}S\left( \mathbf{k}\right) \left[ 1+\frac{2\rho _{s}^{2}\pi
^{2}/A}{1+Z_{V}^{\left( 0\right) }\rho _{s}^{2}\pi ^{2}/A}+\left( 1+f\right)
\left( 1+\frac{\rho _{s}^{2}\pi ^{2}/A}{1+Z_{V}^{\left( 0\right) }\rho
_{s}^{2}\pi ^{2}/A}\right) \right] \\
&=&\phi _{0}^{2}S\left( \mathbf{k}\right) \left( 2+f+\frac{3+f}{A/\rho
_{s}^{2}+Z_{V}^{\left( 0\right) }}\right)
\end{eqnarray*}
The results for Eq.(\ref{eq367}) can now be collected 
\begin{eqnarray}
&&\frac{1}{\phi _{0}^{2}}\left\langle \phi _{V}\phi _{V}\right\rangle _{%
\mathbf{k}}^{vort+cs}  \label{eq64} \\
&=&\frac{1}{k^{2}\rho _{s}^{2}}\left( 1+\frac{1}{k^{2}\rho _{s}^{2}}\right) 
\frac{1}{8\pi ^{2}}\left( 1-\frac{\rho _{s}^{2}k^{2}+1}{\rho
_{s}^{2}k^{2}+1+Z_{V}^{\left( 0\right) }4\pi ^{2}\rho _{s}^{2}/A}\right) 
\notag \\
&&+S\left( \mathbf{k}\right) \left( 2+f+\frac{3+f}{A/\rho
_{s}^{2}+Z_{V}^{\left( 0\right) }}\right)  \notag
\end{eqnarray}
We can make few remarks here. If the arbitrary position in plane of the
vortices and the interaction between physical vortices were neglected, the
only term that would persist is $\exp \left[ Z_{V}^{\left( 0\right) }\right] 
$. The first $\mathbf{k}$-dependent factor in Eq.(\ref{eq54}) comes from
assuming that a statistical ensemble of realizations of the vorticity field
is generated from the random positions in plane of the vortices, even
reduced at a $\delta $-type shape. In practical terms this may be
represented as follows: in a plane, an ensemble of vortices can be placed at
arbitrary positions. We construct the statistical ensemble of the
realizations of this stochastic system. If we measure in one point the
field, it will be zero for most of the realizations and it will be finite
when it happens that a vortex is there. This is a random variable. Now, if
we measure in two points and collect the results for all realizations, the
statistical properties of this quantity (the two-point auto-correlation) has
a Fourier transform that is given by the two factors multiplying the square
bracket in Eq.(\ref{eq54}), divided to $k^{4}$ (since we have the
auto-correlation of the vorticity). When the interaction is considered, the
factor in the curly bracket appears.

\section{Random field influenced by vortices with random positions}

Consider the equation 
\begin{equation}
F\left[ \phi \right] \equiv \nabla _{\perp }^{2}\varphi -\alpha \varphi
-\beta \varphi ^{2}=0  \label{eq65}
\end{equation}
and extract from the total function the part that is due to the vortices 
\begin{equation}
\varphi \left( x,y\right) =\sum_{a=1}^{N}\phi _{s}^{\left( a\right) }\left(
x,y\right) +\phi \left( x,y\right)  \label{eq66}
\end{equation}
Replacing in the equation we have 
\begin{eqnarray}
&&\nabla _{\perp }^{2}\left[ \sum_{a=1}^{N}\phi _{s}^{\left( a\right)
}\left( x,y\right) \right] -\alpha \sum_{a=1}^{N}\phi _{s}^{\left( a\right)
}\left( x,y\right) -\beta \left[ \sum_{a=1}^{N}\phi _{s}^{\left( a\right)
}\left( x,y\right) \right] ^{2}  \label{eq67} \\
&&-2\beta \left[ \sum_{a=1}^{N}\phi _{s}^{\left( a\right) }\left( x,y\right) %
\right] \phi \left( x,y\right)  \notag \\
&&+\nabla _{\perp }^{2}\phi -\alpha \phi -\beta \phi ^{2}  \notag \\
&=&0  \notag
\end{eqnarray}
The first line is zero and we have 
\begin{equation}
\nabla _{\perp }^{2}\phi -\left[ \alpha +2\beta \sum_{a=1}^{N}\phi
_{s}^{\left( a\right) }\left( x,y\right) \right] \phi -\beta \phi ^{2}=0
\label{eq68}
\end{equation}
We write a Lagrangean for the random field according to the MSR procedure 
\begin{equation}
\mathcal{L}\left[ \chi ,\phi \right] =\chi \left\{ \nabla _{\perp }^{2}\phi -%
\left[ \alpha +2\beta \sum_{a=1}^{N}\phi _{s}^{\left( a\right) }\left(
x,y\right) \right] \phi -\beta \phi ^{2}\right\}  \label{eq69}
\end{equation}
and the action functional is 
\begin{eqnarray}
S_{\varphi V}\left[ \chi ,\phi \right] &=&\int dxdy\chi \left\{ \nabla
_{\perp }^{2}\phi -\left[ \alpha +2\beta \sum_{a=1}^{N}\phi _{s}^{\left(
a\right) }\left( x,y\right) \right] \phi -\beta \phi ^{2}\right\}
\label{eq70} \\
&=&\int dxdy\left\{ -\left( \mathbf{\nabla }\chi \right) \left( \mathbf{%
\nabla }\phi \right) -\chi \left[ \alpha +2\beta \sum_{a=1}^{N}\phi
_{s}^{\left( a\right) }\left( x,y\right) \right] \phi -\beta \chi \phi
^{2}\right\}  \notag
\end{eqnarray}
The generating functional is defined from the functional integral 
\begin{equation}
\mathcal{N}^{-1}\int D\left[ \chi \right] D\left[ \phi \right] \exp \left(
-S_{\varphi V}\left[ \chi ,\phi \right] \right)  \label{eq71}
\end{equation}
Now we will modify the action by considering as usual the interaction with
external currents, 
\begin{equation}
\Xi \left[ J\right] \equiv \mathcal{N}^{-1}\int D\left[ \chi \right] D\left[
\phi \right] \exp \left( -S_{\varphi V}\left[ \chi ,\phi \right] +J_{\chi
}\chi +J_{\phi }\phi \right)  \label{eq72}
\end{equation}
With this functional integral we will have to calculate the \emph{free energy%
} functional. The functional $\Xi \left[ J\right] $ depends on the function
representing the vortices. The vortices are assumed known but their position
in plane is random therefore we have to average over them: $-W\left[ J\right]
=\left\langle \ln \left( \Xi \left[ J\right] \right) \right\rangle $.

\subsection{The average over the positions}

To perform the statistical average over the random positions of the
vortices. The functional which we have to average is 
\begin{eqnarray}
\Xi &=&\mathcal{N}^{-1}\int D\left[ \chi \right] D\left[ \phi \right]
\label{eq74} \\
&&\times \exp \left\{ \int dxdy\left[ \left( \mathbf{\nabla }\chi \right)
\left( \mathbf{\nabla }\phi \right) +\alpha \chi \phi +\beta \chi \phi
^{2}\right. \right.  \notag \\
&&\left. \left. +\left( 2\beta \sum_{a=1}^{N}\phi _{s}^{\left( a\right)
}\left( x,y\right) \right) \chi \phi \right] \right\}  \notag
\end{eqnarray}
The part that depends on the positions can be written 
\begin{eqnarray}
&&\left\langle \exp \left\{ \int dxdy\left( 2\beta \sum_{a=1}^{N}\phi
_{s}^{\left( a\right) }\left( x,y\right) \right) \chi \phi \right\}
\right\rangle  \label{eq75} \\
&=&\left\langle \exp \left\{ \int dxdy2\beta \phi ^{s}\sum_{a=1}^{N}\delta
\left( \mathbf{r-r}_{a}\right) \chi \phi \right\} \right\rangle  \notag \\
&=&\left\langle \exp \left[ 2\beta \phi ^{s}\sum_{a=1}^{N}\chi \left( 
\mathbf{r}_{a}\right) \phi \left( \mathbf{r}_{a}\right) \right] \right\rangle
\notag
\end{eqnarray}
where $\phi ^{s}$ is now a simple amplitude. Consider the more general
situation where we have to average in addition over the amplitudes $\phi
^{s} $ of the vortices. If $\phi ^{s}$ is a stochastic variable we have to
perform an average of the type 
\begin{eqnarray}
&&\left\langle \exp \left[ 2\beta \phi ^{s}\sum_{a=1}^{N}\chi \left( \mathbf{%
r}_{a}\right) \phi \left( \mathbf{r}_{a}\right) \right] \right\rangle _{%
\mathbf{r}_{a},\phi ^{s}}  \label{eq76} \\
&=&\prod_{a=1}^{N}\left\langle \exp \left[ 2\beta \phi ^{s}\chi \left( 
\mathbf{r}_{a}\right) \phi \left( \mathbf{r}_{a}\right) \right]
\right\rangle _{\mathbf{r}_{a},\phi ^{s}}  \notag
\end{eqnarray}
Consider that the (now) random variable $\phi ^{s}$ has the probability
density 
\begin{equation}
g\left( \phi ^{s}\right)  \label{eq77}
\end{equation}
Then, restricting for the moment to only the average over $\phi ^{s}$, 
\begin{eqnarray}
&&\left\langle \exp \left[ 2\beta \phi ^{s}\chi \left( \mathbf{r}_{a}\right)
\phi \left( \mathbf{r}_{a}\right) \right] \right\rangle _{\phi ^{s}}
\label{eq78} \\
&=&\int_{-\infty }^{\infty }d\phi ^{s}g\left( \phi ^{s}\right) \exp \left[
2\beta \phi ^{s}\chi \left( \mathbf{r}_{a}\right) \phi \left( \mathbf{r}%
_{a}\right) \right]  \notag \\
&=&\int_{-\infty }^{\infty }d\phi ^{s}g\left( \phi ^{s}\right) \exp \left(
i\lambda \phi ^{s}\right)  \notag
\end{eqnarray}
where 
\begin{equation}
i\lambda \left( \mathbf{r}_{a}\right) \equiv 2\beta \chi \left( \mathbf{r}%
_{a}\right) \phi \left( \mathbf{r}_{a}\right)  \label{eq79}
\end{equation}
Then 
\begin{equation}
\left\langle \exp \left[ 2\beta \phi ^{s}\chi \left( \mathbf{r}_{a}\right)
\phi \left( \mathbf{r}_{a}\right) \right] \right\rangle _{\phi ^{s}}=%
\widetilde{g}\left[ \lambda \left( \mathbf{r}_{a}\right) \right]
\label{eq80}
\end{equation}
where $\widetilde{g}$ is the Fourier transform of the probability
distribution function $g$. We use the notation 
\begin{equation}
\widetilde{g}\left( \mathbf{r}_{k}\right) \equiv \widetilde{g}\left[ \lambda
\left( \mathbf{r}_{a}\right) \right]  \label{eq81}
\end{equation}
and we have to calculate 
\begin{equation}
\prod_{a=1}^{N}\left\langle \widetilde{g}\left( -i2\beta \chi \left( \mathbf{%
r}_{a}\right) \phi \left( \mathbf{r}_{a}\right) \right) \right\rangle _{%
\mathbf{r}_{a}}=\prod_{a=1}^{N}\left\langle \widetilde{g}\left( \mathbf{r}%
_{a}\right) \right\rangle  \label{eq82}
\end{equation}

For this we introduce the function $h$%
\begin{equation}
\widetilde{g}\left( \mathbf{r}_{a}\right) \equiv h\left( \mathbf{r}%
_{a}\right) +1  \label{eq83}
\end{equation}
and we rewrite the average as 
\begin{eqnarray}
&&\prod_{a=1}^{N}\left\langle \widetilde{g}\left( \mathbf{r}_{a}\right)
\right\rangle  \label{eq84} \\
&=&\prod_{a=1}^{N}\left\langle h\left( \mathbf{r}_{a}\right) +1\right\rangle
\notag \\
&=&\left\langle \sum_{l=0}^{\infty }\sum_{i_{1}<i_{2}<...<i_{l}}h\left( 
\mathbf{r}_{i_{1}}\right) h\left( \mathbf{r}_{i_{2}}\right) ...h\left( 
\mathbf{r}_{i_{l}}\right) \right\rangle  \notag \\
&=&\exp \left[ \sum_{l=0}^{\infty }\frac{1}{l!}\int d\mathbf{r}_{i_{1}}d%
\mathbf{r}_{i_{2}}...d\mathbf{r}_{i_{l}}h\left( \mathbf{r}_{i_{1}}\right)
h\left( \mathbf{r}_{i_{2}}\right) ...h\left( \mathbf{r}_{i_{l}}\right)
C^{\left( l\right) }\left( \mathbf{r}_{i_{1}},\mathbf{r}_{i_{2}},...,\mathbf{%
r}_{i_{l}}\right) \right]  \notag
\end{eqnarray}
where we have introduced the cumulants of the distribution of points $%
\mathbf{r}_{i_{1}}$ in the plane.

According to our assumption 
\begin{equation}
C^{\left( l\right) }=\left\{ 
\begin{array}{cc}
1/A & l=1 \\ 
0 & l>1
\end{array}
\right.  \label{eq85}
\end{equation}
where $A$ is the area in plane. Therefore we have the result of averaging 
\begin{eqnarray}
\prod_{a=1}^{N}\left\langle \widetilde{g}\left( \mathbf{r}_{a}\right)
\right\rangle &=&\exp \left[ \frac{1}{A}\int d\mathbf{r}h\left( \mathbf{r}%
\right) \right]  \label{eq86} \\
&=&\exp \left\{ \frac{1}{A}\int d\mathbf{r}\left[ \widetilde{g}\left( 
\mathbf{r}\right) -1\right] \right\}  \notag
\end{eqnarray}
The part of the generating functional which depended on the positions can
now be written 
\begin{equation}
\left\langle \exp \left[ 2\beta \phi ^{s}\sum_{a=1}^{N}\chi \left( \mathbf{r}%
_{a}\right) \phi \left( \mathbf{r}_{a}\right) \right] \right\rangle _{%
\mathbf{r}_{a},\phi ^{s}}=\exp \left\{ \frac{1}{A}\int d\mathbf{r}\left[ 
\widetilde{g}\left( \mathbf{r}\right) -1\right] \right\}  \label{eq87}
\end{equation}
and the full partition function is 
\begin{eqnarray}
\left\langle \Xi \right\rangle _{\mathbf{r}_{a},\phi ^{s}} &=&N^{-1}\int D%
\left[ \chi \right] D\left[ \phi \right]  \label{eq88} \\
&&\times \exp \left\{ \int dxdy\left[ \left( \mathbf{\nabla }\chi \right)
\left( \mathbf{\nabla }\phi \right) +\alpha \chi \phi +\beta \chi \phi
^{2}\right. \right.  \notag \\
&&\left. \left. +\frac{1}{A}\int d\mathbf{r}\left[ \widetilde{g}\left( 
\mathbf{r}\right) -1\right] \right] \right\}  \notag
\end{eqnarray}
The action at the exponent is 
\begin{equation}
\int dxdy\left\{ \left( \mathbf{\nabla }\chi \right) \left( \mathbf{\nabla }%
\phi \right) +\alpha \chi \phi +\beta \chi \phi ^{2}+\frac{1}{A}\left[ 
\widetilde{g}\left( x,y\right) -1\right] \right\}  \label{eq89}
\end{equation}
The function $\widetilde{g}\left( \mathbf{r}\right) $ has as argument the
expression $-i2\beta \chi \left( \mathbf{r}\right) \phi \left( \mathbf{r}%
\right) $.

A reasonable assumption is that the vortices can only be positive or
negative and with the same magnitude. Then we have 
\begin{equation}
g\left( \phi ^{s}\right) =\frac{1}{2}\left[ \delta \left( \phi ^{s}-\phi
_{0}\right) +\delta \left( \phi ^{s}+\phi _{0}\right) \right]  \label{eq90}
\end{equation}
and the Fourier transform is 
\begin{eqnarray}
\widetilde{g}\left( q\right) &=&\int d\phi ^{s}\exp \left( iq\phi
^{s}\right) \frac{1}{2}\left[ \delta \left( \phi ^{s}-\phi _{0}\right)
+\delta \left( \phi ^{s}+\phi _{0}\right) \right]  \label{eq91} \\
&=&\frac{1}{2}\left[ \exp \left( iq\phi _{0}\right) +\exp \left( -iq\phi
_{0}\right) \right]  \notag \\
&=&\cos \left( q\phi _{0}\right)  \notag
\end{eqnarray}
This must be calculated for the argument $-i2\beta \chi \left( \mathbf{r}%
\right) \phi \left( \mathbf{r}\right) $ and gives 
\begin{eqnarray}
\widetilde{g}\left( \mathbf{r}\right) &=&\cos \left[ -i2\beta \phi _{0}\chi
\left( \mathbf{r}\right) \phi \left( \mathbf{r}\right) \right]  \label{eq92}
\\
&=&\cosh \left[ 2\beta \phi _{0}\chi \left( \mathbf{r}\right) \phi \left( 
\mathbf{r}\right) \right]  \notag
\end{eqnarray}
The functional integral that must be calculated becomes 
\begin{eqnarray}
\left\langle \Xi \right\rangle _{\mathbf{r}_{k},\phi ^{s}} &=&\mathcal{N}%
^{-1}\int D\left[ \chi \right] D\left[ \phi \right]  \label{eq93} \\
&&\times \exp \left\{ \int dxdy\left[ \left( \mathbf{\nabla }\chi \right)
\left( \mathbf{\nabla }\phi \right) +\alpha \chi \phi +\beta \chi \phi ^{2}+%
\frac{1}{A}\cosh \left( 2\beta \phi _{0}\chi \phi \right) \right] \right\} 
\notag
\end{eqnarray}
The system is perturbed with an external current $J\left( x,y\right) $
acting on the field $\phi \left( x,y\right) $. 
\begin{eqnarray}
Z_{J} &\equiv &\left\langle \Xi \right\rangle _{\mathbf{r}_{a},\phi ^{s}}%
\left[ J\right]  \label{eq94} \\
&=&\mathcal{N}^{-1}\int D\left[ \chi \right] D\left[ \phi \right] \exp
\left( S_{J}\right)  \notag
\end{eqnarray}
\begin{equation}
S_{J}\equiv \int dxdy\left[ \left( \mathbf{\nabla }\chi \right) \left( 
\mathbf{\nabla }\phi \right) +\alpha \chi \phi +\beta \chi \phi ^{2}+\frac{1%
}{A}\cosh \left( 2\beta \phi _{0}\chi \phi \right) +J\phi \right]
\label{eq95}
\end{equation}

\section{Approximation for small amplitude vortices}

Consider that the amplitudes of the vortices are not high, $\phi _{0}$. Then 
\begin{equation}
\cosh \left( 2\beta \phi _{0}\chi \phi \right) \simeq 1+\frac{1}{2}\left(
2\beta \phi _{0}\chi \phi \right) ^{2}  \label{eq96}
\end{equation}
and the action (removing some terms without significance) 
\begin{equation}
S_{J}\equiv \int dxdy\left[ \left( \mathbf{\nabla }\chi \right) \left( 
\mathbf{\nabla }\phi \right) +\alpha \chi \phi +\beta \chi \phi ^{2}+\frac{%
\left( 2\beta \phi _{0}\right) ^{2}}{A}\chi ^{2}\phi ^{2}+J\phi \right]
\label{eq97}
\end{equation}

This action in principle can lead to a perturbative treatment but with two
vertices, of order three and of order four, a very difficult and unusual
problem. We remark however that 
\begin{equation}
S_{J}\equiv \int dxdy\left[ -\chi \left( \mathbf{\nabla }^{2}\phi \right)
+\alpha \chi \phi +\beta \chi \phi ^{2}+\frac{\left( 2\beta \phi _{0}\right)
^{2}}{A}\chi ^{2}\phi ^{2}+J\phi \right]  \label{eq98}
\end{equation}
may become quadratic in $\chi $ and in $\phi $ (therefore the functional
integral is Gaussian) if we succeed to separate the product $\chi ^{2}\phi
^{2}$.

\subsubsection{Technical step}

Consider the following formula to disentangle the two variables 
\begin{equation}
\exp \left( \frac{1}{2}U^{2}\right) =\frac{1}{\sqrt{2\pi }}\int_{-\infty
}^{\infty }ds\exp \left( -\frac{1}{2}s^{2}-Us\right)  \label{eq99}
\end{equation}
then 
\begin{equation}
\exp \left( \frac{1}{2}2Q\chi ^{2}\phi ^{2}\right) =\frac{1}{\sqrt{2\pi }}%
\int_{-\infty }^{\infty }ds\exp \left( -\frac{1}{2}s^{2}-\sqrt{2Q}\chi \phi
s\right)  \label{eq100}
\end{equation}
and we have the action 
\begin{eqnarray}
Z_{J} &=&\mathcal{N}^{-1}\frac{1}{\sqrt{2\pi }}\int_{-\infty }^{\infty
}ds\exp \left( -\frac{1}{2}s^{2}\right)  \label{eq101} \\
&&\hspace*{-0.5cm}\times \int D\left[ \phi \right] D\left[ \chi \right] \exp
\left\{ \int dxdy\left[ -\chi \left( \mathbf{\nabla }^{2}\phi \right)
+\alpha \chi \phi +\beta \chi \phi ^{2}-\sqrt{2Q}\chi \phi s+J\phi \right]
\right\}  \notag
\end{eqnarray}
or 
\begin{equation}
S_{J}=\int dxdy\left[ J\phi \right] +\int dxdy\left[ -\chi \left( \mathbf{%
\nabla }^{2}\phi \right) +\chi \left( \alpha -s\sqrt{2Q}\right) \phi +\beta
\chi \phi ^{2}\right]  \label{eq102}
\end{equation}
We have used the notation 
\begin{equation}
Q\equiv \frac{\left( 2\beta \phi _{0}\right) ^{2}}{A}  \label{eq103}
\end{equation}
The functional integration over $\chi $ can be done immediately and gives a
functional $\delta $ with the argument the equation in the modified form. If
we make the integration over $\chi $ we obtain 
\begin{eqnarray}
\hspace*{-0.5cm} &&\int D\left[ \chi \right] \exp \left\{ \int dxdy\left[
-\chi \left( \mathbf{\nabla }^{2}\phi \right) +\chi \left( \alpha -s\sqrt{2Q}%
\right) \phi +\beta \chi \phi ^{2}\right] \right\}  \label{eq104} \\
&=&\int D\left[ \chi \right] \exp \left[ \int dxdy\chi \overline{F}\left(
\phi \right) \right]  \notag \\
&=&\left[ \left| -i\frac{\delta \overline{F}}{\delta \phi }\right| _{\phi
^{z}}\right] ^{-1}\delta \left( \phi -\phi ^{z}\right)  \notag
\end{eqnarray}
where we have introduced the notation 
\begin{equation}
\overline{F}\left( \phi \right) \equiv -\mathbf{\nabla }^{2}\phi +\left(
\alpha -s\sqrt{2Q}\right) \phi +\beta \phi ^{2}  \label{eq105}
\end{equation}
and the function $\overline{\phi }^{z}$ is the solution of the differential
equation $\overline{F}\left( \phi \right) =0$ \emph{i.e.} 
\begin{equation}
\overline{F}\left( \overline{\phi }^{z}\right) =0  \label{eq106}
\end{equation}

Inserting this result in the integral over the functions $\phi $, we have 
\begin{eqnarray}
&&\mathcal{N}^{-1}\int D\left[ \phi \right] \exp \left( \int dxdyJ\phi
\right) \left[ \left| -i\frac{\delta \overline{F}}{\delta \phi }\right| _{%
\overline{\phi }^{z}}\right] ^{-1}\delta \left( \phi -\overline{\phi }%
^{z}\right)  \label{eq107} \\
&=&\left[ \left| -i\frac{\delta \overline{F}}{\delta \phi }\right| _{%
\overline{\phi }^{z}}\right] ^{-1}\exp \left( \int dxdyJ\overline{\phi }%
^{z}\right)  \notag
\end{eqnarray}
Then 
\begin{equation}
Z_{J}=\mathcal{N}^{-1}\frac{1}{\sqrt{2\pi }}\int_{-\infty }^{\infty }ds\exp
\left( -\frac{1}{2}s^{2}\right) \left[ \left| -i\frac{\delta \overline{F}}{%
\delta \phi }\right| _{\overline{\phi }^{z}}\right] ^{-1}\exp \left( \int
dxdyJ\overline{\phi }^{z}\right)  \label{eq108}
\end{equation}

The most important part is to calculate 
\begin{equation}
-i\left. \frac{\delta \overline{F}}{\delta \phi }\right| _{\overline{\phi }%
^{z}}=-i\left. \frac{\delta }{\delta \phi }\left[ -\mathbf{\nabla }^{2}\phi
+\left( \alpha -s\sqrt{2Q}\right) \phi +\beta \phi ^{2}\right] \right| _{%
\overline{\phi }^{z}}  \label{eq109}
\end{equation}
But this simplifies with the part of the factor 
\begin{equation}
\mathcal{N}^{-1}\sim \left| \frac{\delta F}{\delta \phi }\right| _{\phi ^{z}}
\label{eq110}
\end{equation}
The fact that the initial factor is written before averaging over the
positions and the second factor is obtained after averaging will make a
small difference which can only be of higher order. What remains is a factor
of normalization that we call $c^{-1}$.

The functional derivatives are 
\begin{equation}
\frac{\delta Z_{j}}{\delta J\left( x,y\right) }=\frac{1}{c}\frac{1}{\sqrt{%
2\pi }}\int_{-\infty }^{\infty }ds\exp \left( -\frac{1}{2}s^{2}\right) 
\overline{\phi }^{z}\left( x,y\right) \exp \left( \int dxdyJ\overline{\phi }%
^{z}\right)  \label{eq111}
\end{equation}
and 
\begin{eqnarray}
\frac{\delta ^{2}Z_{J}}{\delta J\left( x,y\right) \delta J\left( x^{\prime
},y^{\prime }\right) } &=&\frac{1}{c}\frac{1}{\sqrt{2\pi }}\int_{-\infty
}^{\infty }ds\exp \left( -\frac{1}{2}s^{2}\right)  \label{eq112} \\
&&\times \overline{\phi }^{z}\left( x,y\right) \overline{\phi }^{z}\left(
x^{\prime },y^{\prime }\right) \exp \left( \int dxdyJ\overline{\phi }%
^{z}\right)  \notag
\end{eqnarray}
and taking 
\begin{equation}
J\equiv 0  \label{eq113}
\end{equation}
we can calculate the contribution to the two-point correlation arising from
this part of the partition function.

We have to solve 
\begin{equation}
-\mathbf{\nabla }^{2}\phi +\left( \alpha -s\sqrt{2Q}\right) \phi +\beta \phi
^{2}=0  \label{eq114}
\end{equation}
for a small amplitude field $\phi $. This can be done by approximations,
starting from the solution of the equation without random centers.

We can see that the immediate effect of the scattering of the turbulent
field by the vortices is a renormalization of the coefficient $\alpha $ of
the equation.

The solution that we need for Eq.(\ref{eq114}) must reflect the fact that we
are examining a turbulent field, in the dynamics of which the randomly
placed vortices have been included. We can see that the integration over the
auxiliary variable $s$ must not necessarily be truncated to ensure that $%
\alpha _{s}\equiv \alpha -s\sqrt{2Q}$ remains positive. This is because the
field is turbulent and there is no restriction concerning the existence of
the vortex solution which should be imposed to the turbulent component of
the field. We also note that the approximation taking only the $s=0$
contribution in the integral (\ref{eq112}) returns us to the problem of the
turbulent field without interaction with the random vortices. Although the
main contribution (as shown by the Gaussian integration) comes from the free
turbulent field itself, the presence of vortices is contained in the rest of
the integral, for $\left| s\right| $ not close to $0$.

These remarks suggest to study the statistical properties of the transformed
field, where $\alpha $ is replaced with $\alpha _{s}$.

\section{The background turbulence: perturbative treatment}

The vortices represent the strongly nonlinear part of the system and they
have been extracted and treated separately as shown in Section IV. However,
there is still nonlinear interaction in the remaining random field. This is
the nonlinear mode coupling which produces the stationary turbulent states
of the system, even in absence of any definite structure formation. This
interaction must be taken into account when we analyse the statistical
properties of the turbulent field. From general consideration we know what
can be expected from this analysis: we will obtain a small departure from
pure Gaussian statistics, expressed in nonlinear renormalization of the
propagator and of the vertex of the interaction. The presence of random
interaction with elements of the system that are beyond our simple model is
accounted, as usual, by a noise term acting like a drive for the system of
random waves. The noise will be assumed with the simplest (white) statistics 
\begin{equation}
\left\langle \zeta \left( x,y\right) \zeta \left( x^{\prime },y^{\prime
}\right) \right\rangle =D\delta \left( x-x^{\prime }\right) \delta \left(
y-y^{\prime }\right)  \label{eq120}
\end{equation}
and should be considered as a random stirring force composed, for example,
of random growths and decays of marginally stable modes, thus injecting at
random places some energy into the system.

We will provide a standard perturbative treatment for the differential
equation 
\begin{equation}
-\mathbf{\nabla }^{2}\phi +\left( \alpha -s\sqrt{2A}\right) \phi +\beta \phi
^{2}=\zeta  \label{eq121}
\end{equation}
with the objective to calculate correlation functions and other statistical
properties.

As usual we start by defining the generating functional of the statistical
correlation 
\begin{eqnarray}
&&\exp \left( -W\left[ \mathbf{J}\right] \right)  \label{eq122} \\
&=&\int \mathcal{D}\left[ \eta \right] \mathcal{D}\left[ \phi \right] \exp
\left\{ \int dxdy\left[ -\eta \mathbf{\nabla }^{2}\phi +\eta \alpha _{s}\phi
+\eta \beta \phi ^{2}+J_{\eta }\eta +J_{\phi }\phi \right] \right\}  \notag
\\
&&\times \left\langle \int \mathcal{D}\left[ \eta \right] \mathcal{D}\left[
\phi \right] \exp \left\{ \int dxdy\eta \zeta \right\} \right\rangle  \notag
\end{eqnarray}
where we have introduced the notation 
\begin{equation}
\alpha _{s}\equiv \alpha -s\sqrt{2Q}  \label{eq123}
\end{equation}
The second factor can easily be calculated 
\begin{eqnarray}
&&\left\langle \int \mathcal{D}\left[ \eta \right] \mathcal{D}\left[ \phi %
\right] \exp \left( \int dxdy\eta \zeta \right) \right\rangle  \label{eq124}
\\
&=&\int \mathcal{D}\left[ \eta \right] \mathcal{D}\left[ \phi \right] \exp %
\left[ \int dxdy\left( 2D\eta ^{2}\right) \right]  \notag
\end{eqnarray}
In the generating functional there are the two dual functions, $\eta \left(
x,y\right) $ and $\phi \left( x,y\right) $, the second being the physical
field of the random waves. We have inserted the action related with
interaction with ``external'' currents, $J_{\eta }$ and $J_{\phi }$ which
will permit to express the averages as functional derivatives.

Defining the Lagrangean density (after an integration by parts) 
\begin{equation}
\mathcal{L}=\left( \mathbf{\nabla }\eta \right) \left( \mathbf{\nabla }\phi
\right) +\eta \alpha _{s}\phi +\eta \beta \phi ^{2}+D\eta ^{2}+J_{\eta }\eta
+J_{\phi }\phi  \label{eq125}
\end{equation}
we will separate into the Gaussian the nonlinear interaction part 
\begin{equation}
\mathcal{L}=\mathcal{L}_{0}+\mathcal{L}_{I}  \label{eq126}
\end{equation}
\begin{equation}
\mathcal{L}_{0}=\left( \mathbf{\nabla }\eta \right) \left( \mathbf{\nabla }%
\phi \right) +\eta \alpha _{s}\phi +D\eta ^{2}+J_{\eta }\eta +J_{\phi }\phi
\label{eq127}
\end{equation}
\begin{equation}
\mathcal{L}_{I}=\eta \beta \phi ^{2}  \label{eq128}
\end{equation}
The first part of the Lagrangean is not linear because of the square term,
but it is Gaussian therefore it would not pose any problem to the functional
integration. We note that we have an order three vertex.

The action functional is correspondingly divided into two parts and can be
written, as usual 
\begin{eqnarray}
Z\left[ \mathbf{J}\right] &=&\int \mathcal{D}\left[ \eta \right] \mathcal{D}%
\left[ \phi \right] \exp \left( \int dxdy\mathcal{L}_{0}\right) \exp \left(
\int dxdy\mathcal{L}_{I}\right)  \label{eq129} \\
&=&\exp \left( \beta \int dxdy\frac{\delta }{\delta J_{\eta }}\frac{\delta }{%
\delta J_{\phi }}\frac{\delta }{\delta J_{\phi }}\right) \int \mathcal{D}%
\left[ \eta \right] \mathcal{D}\left[ \phi \right] \exp \left( \int dxdy%
\mathcal{L}_{0}\right)  \notag
\end{eqnarray}
For the linear part we have the Euler-Lagrange equations 
\begin{equation}
\frac{d}{dx}\frac{\delta \mathcal{L}_{0}}{\delta \left( \partial \eta
/\partial x\right) }+\frac{d}{dy}\frac{\delta \mathcal{L}_{0}}{\delta \left(
\partial \eta /\partial y\right) }-\frac{\delta \mathcal{L}_{0}}{\delta \eta 
}=0  \label{eq130}
\end{equation}
\begin{equation}
\frac{d}{dx}\frac{\delta \mathcal{L}_{0}}{\delta \left( \partial \phi
/\partial x\right) }+\frac{d}{dy}\frac{\delta \mathcal{L}_{0}}{\delta \left(
\partial \phi /\partial y\right) }-\frac{\delta \mathcal{L}_{0}}{\delta \phi 
}=0  \label{eq131}
\end{equation}
The equations are 
\begin{equation}
\Delta \phi -\alpha _{s}\phi -2D\eta =J_{\eta }  \label{eq132}
\end{equation}
and 
\begin{equation}
\Delta \eta -\alpha _{s}\eta =J_{\phi }  \label{eq133}
\end{equation}
The latter is an inhomogeneous Helmholtz equation and has the solution 
\begin{equation}
\eta _{0}\left( x,y\right) =\int dx^{\prime }dy^{\prime }G_{\eta \phi
}\left( x,y;x^{\prime },y^{\prime }\right) J_{\phi }\left( x^{\prime
},y^{\prime }\right)  \label{eq134}
\end{equation}
in terms of the Green function appropriate for the space domain of our
analysis and taking into account the boundary conditions for $\eta $.

The first Euler-Lagrange equation gives 
\begin{eqnarray}
\Delta \phi -\alpha _{s}\phi &=&J_{\eta }+2D\eta _{0}  \label{eq135} \\
&=&J_{\eta }+2D\int dx^{\prime }dy^{\prime }G_{\eta \phi }\left(
x,y;x^{\prime },y^{\prime }\right) J_{\phi }\left( x^{\prime },y^{\prime
}\right)  \notag
\end{eqnarray}
with the solution 
\begin{eqnarray}
\phi _{0}\left( x,y\right) &=&\int dx^{\prime }dy^{\prime }G_{\phi \eta
}\left( x,y;x^{\prime },y^{\prime }\right) J_{\eta }\left( x^{\prime
},y^{\prime }\right)  \label{eq136} \\
&&+2D\int dx^{\prime }dy^{\prime }dx^{\prime \prime }dy^{\prime \prime
}G_{\phi \eta }\left( x,y;x^{\prime \prime },y^{\prime \prime }\right)
G_{\eta \phi }\left( x^{\prime \prime },y^{\prime \prime };x^{\prime
},y^{\prime }\right) J_{\phi }\left( x^{\prime },y^{\prime }\right)  \notag
\end{eqnarray}
We can introduce the matrix of propagators 
\begin{equation}
\mathbf{G\equiv }\left( G_{ij}\right) =\left( 
\begin{array}{cc}
G_{\phi \phi } & G_{\phi \eta } \\ 
G_{\eta \phi } & 0
\end{array}
\right)  \label{eq137}
\end{equation}
where 
\begin{equation}
G_{\phi \phi }\left( x,y;x^{\prime },y^{\prime }\right) =2D\int dx^{\prime
\prime }dy^{\prime \prime }G_{\phi \eta }\left( x,y;x^{\prime \prime
},y^{\prime \prime }\right) G_{\eta \phi }\left( x^{\prime \prime
},y^{\prime \prime };x^{\prime },y^{\prime }\right)  \label{eq138}
\end{equation}
Introducing the notation 
\begin{equation}
\Psi \equiv \left( 
\begin{array}{c}
\phi \\ 
\eta
\end{array}
\right)  \label{eq139}
\end{equation}
we will write symbolically (summation over repeated indices is implied) 
\begin{equation}
\Psi _{i}=\int dx^{\prime }dy^{\prime }G_{ik}J_{k}  \label{eq140}
\end{equation}
for 
\begin{equation}
\Psi _{i}=\int dx^{\prime }dy^{\prime }G\left( x,y;x^{\prime },y^{\prime
}\right) J_{k}\left( x^{\prime },y^{\prime }\right)  \label{eq141}
\end{equation}

We can now calculate the linear part of the action along the system 's
configurations given by these solutions. The action will be, of course, a
functional of the current. 
\begin{eqnarray}
S_{0J} &=&\int dxdy\left[ \left( \mathbf{\nabla }\eta _{0}\right) \left( 
\mathbf{\nabla }\phi _{0}\right) +\alpha _{s}\eta _{0}\phi _{0}+D\eta
_{0}^{2}+J_{\eta }\eta _{0}+J_{\phi }\phi _{0}\right]  \label{eq142} \\
&=&\frac{1}{2}\int dxdy\left[ -\eta _{0}\Delta \phi _{0}+\alpha _{s}\eta
_{0}\phi _{0}+2D\eta _{0}^{2}+J_{\eta }\eta _{0}+J_{\phi }\phi _{0}\right] 
\notag \\
&&+\frac{1}{2}\int dxdy\left[ -\phi _{0}\Delta \eta _{0}+\alpha _{s}\eta
_{0}\phi _{0}+J_{\eta }\eta _{0}+J_{\phi }\phi _{0}\right]  \notag
\end{eqnarray}
or 
\begin{eqnarray}
S_{0J} &=&\frac{1}{2}\int dxdy\left[ \eta _{0}\left( -\Delta \phi
_{0}+\alpha _{s}\phi _{0}+2D\eta _{0}+J_{\eta }\right) +J_{\phi }\phi _{0}%
\right]  \label{eq143} \\
&&+\frac{1}{2}\int dxdy\left[ \phi _{0}\left( -\Delta \eta _{0}+\alpha
_{s}\eta _{0}+J_{\phi }\right) +J_{\eta }\eta _{0}\right]  \notag \\
&=&\frac{1}{2}\int dxdy\left( J_{\phi }\phi _{0}+J_{\eta }\eta _{0}\right) 
\notag
\end{eqnarray}
We now express the solution by the Green functions 
\begin{eqnarray}
S_{0J} &=&\frac{1}{2}\int dxdy\left[ J_{\phi }\int dx^{\prime }dy^{\prime
}G_{\phi \eta }\left( x,y;x^{\prime },y^{\prime }\right) J_{\eta }\left(
x^{\prime },y^{\prime }\right) \right.  \label{eq144} \\
&&\left. +J_{\eta }\int dx^{\prime }dy^{\prime }G_{\eta \phi }\left(
x,y;x^{\prime },y^{\prime }\right) J_{\phi }\left( x^{\prime },y^{\prime
}\right) \right]  \notag
\end{eqnarray}
which can be written using the convention introduced above 
\begin{equation}
S_{0J}=\frac{1}{2}\int dxdy\int dx^{\prime }dy^{\prime }J_{i}\left(
x,y\right) G_{ij}\left( x,y;x^{\prime },y^{\prime }\right) J_{k}\left(
x^{\prime },y^{\prime }\right)  \label{eq145}
\end{equation}

We return to the generating functional 
\begin{eqnarray}
Z\left[ \mathbf{J}\right] &=&\exp \left( \beta \int dxdy\frac{\delta }{%
\delta J_{\eta }}\frac{\delta }{\delta J_{\phi }}\frac{\delta }{\delta
J_{\phi }}\right) \exp \left( S_{0J}\right)  \label{eq146} \\
&=&\exp \left( \int dxdyC_{ijk}\frac{\delta }{\delta J_{i}}\frac{\delta }{%
\delta J_{j}}\frac{\delta }{\delta J_{k}}\right)  \notag \\
&&\times \exp \left[ \frac{1}{2}\int dxdy\int dx^{\prime }dy^{\prime
}J_{i}\left( x,y\right) G_{ij}\left( x,y;x^{\prime },y^{\prime }\right)
J_{k}\left( x^{\prime },y^{\prime }\right) \right]  \notag
\end{eqnarray}
where we have introduced, for uniformity of notation 
\begin{equation}
C_{ijk}\left( x,y\right) \equiv \beta \delta _{i\eta }\delta _{j\phi }\delta
_{k\phi }  \label{eq147}
\end{equation}
This is a classical framework for diagrammatic expansion.

In general the functional approach is well suited for statistical problems
where the statistical ensemble of the realization of the system's
configurations is produced by the effect of a noise or by a random choice of
initial condition, as explained in MSR and Jensen. In the most usual case,
where there is a noise acting on the system, using the (known) statistical
properties of the noise means implicitly that we make a change of variables,
from the field $\phi $ to the noise $\zeta $ and this introduces a Jacobian, 
\begin{equation}
\mathcal{J}\left[ \phi \right] =\mathcal{D}\left[ \zeta \right] /\mathcal{D}%
\left[ \phi \right]  \label{eq148}
\end{equation}
The Jacobian cancels all the diagrams that correspond to the \emph{vacuum}
to \emph{vacuum} transitions, or equal-coordinates correlations. For this
reason it is not mentioned. This is explained in Appendix B.

\bigskip

When applying the operator to the second exponential we have to remind that
we need at least two open ends with currents $J_{\phi }$ since we intend to
determine the spectrum. On the other hand, it is clear that any open end
with isolated either $J_{\phi }$ or $J_{\eta }$ would vanish after taking
the currents to zero.

The first term is naturally the diffusion driven by the random rise and
decay of marginally stable modes, represented here as a noise.

The next term that is useful consists of the product of two vertex-like
operators applied on a product of four propagator-like terms 
\begin{equation}
J_{i_{1}}G_{i_{1}i_{2}}C_{i_{2}i_{3}i_{4}}G_{i_{4}i_{5}}G_{i_{3}i_{6}}C_{i_{5}i_{6}i_{7}}G_{i_{7}i_{1}}J_{i_{1}}
\label{eq149}
\end{equation}
Obviously, this is the \emph{one-loop} diagram. We will derivate to the
first and last factors (currents) so that, more explicitly, its structure is 
\begin{equation}
G_{\phi \eta }C_{\eta \phi \phi }G_{\phi \phi }G_{\phi \phi }C_{\phi \phi
\eta }G_{\eta \phi }  \label{eq150}
\end{equation}

We can represent the Green function by its Fourier transform 
\begin{equation}
\left( \Delta -\alpha _{s}\right) G\left( x,y;x^{\prime },y^{\prime }\right)
=-\delta \left( x-x^{\prime }\right) \delta \left( y-y^{\prime }\right)
\label{eq151}
\end{equation}
\begin{eqnarray}
G\left( x,y;x^{\prime },y^{\prime }\right) &=&\int dk_{x}dk_{y}\frac{1}{%
k_{x}^{2}+k_{y}^{2}+\alpha _{s}}  \label{eq152} \\
&&\times \exp \left[ -ik_{x}\left( x-x^{\prime }\right) \right] \exp \left[
-ik_{y}\left( y-y^{\prime }\right) \right]  \notag
\end{eqnarray}
We obtain the propagators for the coupling between the field $\phi $ and its
dual $\eta $%
\begin{eqnarray}
G_{\phi \eta }\left( x,y;x^{\prime },y^{\prime }\right) &=&\int dk_{x}dk_{y}%
\frac{1}{k_{x}^{2}+k_{y}^{2}+\alpha _{s}}  \label{eq153} \\
&&\times \exp \left[ -ik_{x}\left( x-x^{\prime }\right) \right] \exp \left[
-ik_{y}\left( y-y^{\prime }\right) \right]  \notag \\
&=&G_{\eta \phi }\left( x,y;x^{\prime },y^{\prime }\right)  \notag
\end{eqnarray}
\begin{eqnarray}
G_{\phi \phi }\left( x,y\right) &=&2D\int dx^{\prime \prime }dy^{\prime
\prime }G_{\phi \eta }\left( x,y;x^{\prime \prime },y^{\prime \prime
}\right) G_{\eta \phi }\left( x^{\prime \prime },y^{\prime \prime
};x^{\prime },y^{\prime }\right)  \label{eq154} \\
&=&\int dk_{x}dk_{y}\frac{1}{\left( k_{x}^{2}+k_{y}^{2}+\alpha _{s}\right)
^{2}}  \notag \\
&&\times \exp \left[ -ik_{x}\left( x-x^{\prime }\right) \right] \exp \left[
-ik_{y}\left( y-y^{\prime }\right) \right]  \notag
\end{eqnarray}
The intermediate integration over space $\left( x^{\prime \prime },y^{\prime
\prime }\right) $ yields equality of the Fourier variables of the two Green
functions. We then have 
\begin{eqnarray}
\widetilde{G}_{\phi \phi } &=&2D\left( k_{x}^{2}+k_{y}^{2}+\alpha
_{s}\right) ^{-2}  \label{eq155} \\
\widetilde{G}_{\phi \eta } &=&\widetilde{G}_{\eta \phi }=\left(
k_{x}^{2}+k_{y}^{2}+\alpha _{s}\right) ^{-1}  \notag
\end{eqnarray}

The one-loop term is expressed as two integrals over the intermediate
momenta at the two vertices where we have a product of three Green functions
with the vertex. Conservation of momentum in the loop and overall
conservation of the diagram (which means that the $\mathbf{k}$ at input line
must be the same at the output line) lead to 
\begin{eqnarray}
&&\left\langle \phi \left( x,y\right) \phi \left( x^{\prime },y^{\prime
}\right) \right\rangle  \label{eq156} \\
&=&\int dk_{x}dk_{y}\left\langle \phi \phi \right\rangle _{\mathbf{k}}\exp %
\left[ -ik_{x}\left( x-x^{\prime }\right) -ik_{y}\left( y-y^{\prime }\right) %
\right]  \notag \\
&=&\frac{2D}{\left( \mathbf{k}^{2}+\alpha _{s}\right) ^{2}}  \notag \\
&&+\int d\mathbf{k}\Gamma \int d\mathbf{p}\frac{1}{\mathbf{k}^{2}+\alpha _{s}%
}\beta \frac{2D}{\left( \mathbf{p}^{2}+\alpha _{s}\right) ^{2}}\frac{2D}{%
\left[ \left( \mathbf{k-p}\right) ^{2}+\alpha _{s}\right] ^{2}}\beta \frac{1%
}{\mathbf{k}^{2}+\alpha _{s}}  \notag
\end{eqnarray}
The coefficient $\Gamma $ represents the multiplicity of this diagram and
factors of normalization. 
\begin{eqnarray}
&&\left\langle \phi \phi \right\rangle _{\mathbf{k}}  \label{eq157} \\
&=&\Gamma \left( 2D\beta \right) ^{2}\left( \frac{1}{\mathbf{k}^{2}+\alpha
_{s}}\right) ^{2}\int d\mathbf{p}\frac{1}{\left( \mathbf{p}^{2}+\alpha
_{s}\right) ^{2}}\frac{1}{\left[ \left( \mathbf{k-p}\right) ^{2}+\alpha _{s}%
\right] ^{2}}  \notag \\
&=&\Gamma \left( 2D\beta \right) ^{2}\left( \frac{1}{\mathbf{k}^{2}+\alpha
_{s}}\right) ^{2}\int d\mathbf{p}\frac{1}{\left[ \left( \mathbf{k/}2-\mathbf{%
p}\right) ^{2}+\alpha _{s}\right] ^{2}}\frac{1}{\left[ \left( \mathbf{k/2}+%
\mathbf{p}\right) ^{2}+\alpha _{s}\right] ^{2}}  \notag
\end{eqnarray}

\bigskip

We transform the integral 
\begin{eqnarray}
I &\equiv &\int pdpd\theta \frac{1}{\left[ \mathbf{p}^{2}-\mathbf{k\cdot p+k}%
^{2}/4+\alpha _{s}\right] ^{2}}  \label{eq158} \\
&&\times \frac{1}{\left[ \mathbf{p}^{2}+\mathbf{k\cdot p+k}^{2}/4+\alpha _{s}%
\right] ^{2}}  \notag
\end{eqnarray}
Now we introduce the variables that will replace the constant terms 
\begin{equation}
b\equiv \mathbf{k}^{2}/4+\alpha _{s}  \label{eq159}
\end{equation}
and 
\begin{eqnarray}
\mathbf{k\cdot p} &=&kp\cos \theta  \label{eq160} \\
&\equiv &up  \notag
\end{eqnarray}
The integral becomes 
\begin{equation}
I=\int pdpd\theta \frac{1}{\left[ p^{2}-up+b\right] ^{2}}\frac{1}{\left[
p^{2}+up+b\right] ^{2}}  \label{eq161}
\end{equation}
We write the denominator in the form 
\begin{eqnarray}
&&\left[ p^{2}-up+b\right] ^{2}\left[ p^{2}+up+b\right] ^{2}  \label{eq162}
\\
&=&\left[ \left( p^{2}+b\right) ^{2}-u^{2}p^{2}\right] ^{2}  \notag \\
&=&\left[ p^{4}+2p^{2}b+b^{2}-u^{2}p^{2}\right] ^{2}  \notag \\
&=&\left[ p^{4}+\left( 2b-u^{2}\right) p^{2}+b^{2}\right] ^{2}  \notag \\
&=&\left( p^{4}+2\gamma p^{2}+\delta \right) ^{2}  \notag
\end{eqnarray}
where 
\begin{eqnarray}
2\gamma &\equiv &2b-u^{2}  \label{eq163} \\
\delta &\equiv &b^{2}  \notag
\end{eqnarray}

The integral becomes 
\begin{equation}
I=\int d\theta \int_{0}^{\infty }\frac{pdp}{\left( p^{4}+2\gamma
p^{2}+\delta \right) ^{2}}  \label{eq164}
\end{equation}
we make the substitution 
\begin{equation}
p^{2}\rightarrow x  \label{eq165}
\end{equation}
\begin{equation}
I=\int d\theta \frac{1}{2}\int_{0}^{\infty }\frac{dx}{\left( x^{2}+2\gamma
x+\delta \right) ^{2}}  \label{eq166}
\end{equation}
The integration over the intermediate momentum is 
\begin{eqnarray}
&&\int_{0}^{\infty }\frac{dx}{\left( x^{2}+2\gamma x+\delta \right) ^{2}}
\label{eq167} \\
&=&\frac{1}{2}\frac{1}{\left( \delta -\gamma ^{2}\right) ^{3/2}}\left[ \frac{%
\pi }{2}-\arctan \frac{\gamma }{\sqrt{\delta -\gamma ^{2}}}\right] +\left( -%
\frac{\gamma }{2\delta }\right) \frac{1}{\delta -\gamma ^{2}}  \notag
\end{eqnarray}
Here 
\begin{eqnarray}
\gamma &=&b-u^{2}/2  \label{eq168} \\
&=&\frac{1}{4}\left[ k^{2}\left( 1-2\cos ^{2}\theta \right) +4\alpha _{s}%
\right]  \notag
\end{eqnarray}
\begin{eqnarray}
\frac{\gamma }{\sqrt{\delta -\gamma ^{2}}} &=&\frac{b-u^{2}/2}{\left|
u\right| \sqrt{b-u^{2}/4}}  \label{eq169} \\
&=&\frac{1}{2}\frac{k^{2}\left( 1-2\cos ^{2}\theta \right) +4\alpha _{s}}{%
k\left| \cos \theta \right| \sqrt{k^{2}\sin ^{2}\theta +4\alpha _{s}}} 
\notag
\end{eqnarray}
\begin{eqnarray}
\delta -\gamma ^{2} &=&u^{2}\left( b-\frac{u^{2}}{4}\right)  \label{eq170} \\
&=&\frac{1}{4}k^{2}\cos ^{2}\theta \left( k^{2}\sin ^{2}\theta +\alpha
_{s}\right)  \notag
\end{eqnarray}
\begin{equation}
\frac{\gamma }{\delta }=4\frac{k^{2}\left( 1-2\cos ^{2}\theta \right)
+4\alpha _{s}}{\left( k^{2}+4\alpha _{s}\right) ^{2}}  \label{eq171}
\end{equation}

Then 
\begin{eqnarray}
&&\int pdpd\theta \frac{1}{\left[ \mathbf{p}^{2}-\mathbf{k\cdot p+k}%
^{2}/4+\alpha _{s}\right] ^{2}}\frac{1}{\left[ \mathbf{p}^{2}+\mathbf{k\cdot
p+k}^{2}/4+\alpha _{s}\right] ^{2}}  \label{eq172} \\
&=&\int d\theta \left\{ \frac{8}{\left[ k^{2}\cos ^{2}\theta \left(
k^{2}\sin ^{2}\theta +\alpha _{s}\right) \right] ^{3/2}}\right.  \notag \\
&&\times \left[ \frac{\pi }{2}-\arctan \frac{k^{2}\left( 1-2\cos ^{2}\theta
\right) +4\alpha _{s}}{2k\left| \cos \theta \right| \sqrt{k^{2}\sin
^{2}\theta +4\alpha _{s}}}\right]  \notag \\
&&\left. -\frac{8}{k^{2}\cos ^{2}\theta \left( k^{2}\sin ^{2}\theta +\alpha
_{s}\right) }\frac{k^{2}\left( 1-2\cos ^{2}\theta \right) +4\alpha _{s}}{%
\left( k^{2}+4\alpha _{s}\right) ^{2}}\right\}  \notag
\end{eqnarray}
We can show that the integral is not singular at the limit 
\begin{equation}
\cos \theta \rightarrow 0  \label{eq173}
\end{equation}
so that the integration over $\theta $ is safe.

As order of magnitude the integral is dominated by terms like 
\begin{equation}
\sim \Gamma \left( 2D\beta \right) ^{2}\frac{1}{k^{2}\left( k^{2}+\alpha
_{s}\right) ^{3/2}}  \label{eq174}
\end{equation}
while the diffusive part is 
\begin{equation}
\sim \frac{2D}{\left( k^{2}+\alpha _{s}\right) ^{2}}  \label{eq175}
\end{equation}

\bigskip

After this we obtain for the turbulent background field 
\begin{equation}
\left\langle \phi \phi \right\rangle _{\mathbf{k}}^{turbulence}=a\frac{2D}{%
\left( k^{2}+\alpha _{s}\right) ^{2}}+b\left( 2D\beta \right) ^{2}\frac{1}{%
\left( k^{2}+\alpha _{s}\right) ^{2}}\frac{1}{k^{2}\left( k^{2}+\alpha
_{s}\right) ^{3/2}}  \label{eq176}
\end{equation}
where we have collected the factors in two numbers, $a$ and $b$.

\section{Summary}

It is usually assumed that a turbulent plasma (in prticular with embedded
structures) should exhibit a spectrum of exponential or algebraic type.
There is no universal theoretical basis for such an assumption except for
cases where the scaling invariance is justified on physical ground. We find
that it would be more adequate to extract a particular behavior (on
different spectral intervals) from expressions like Eq.(\ref{eq64}) and Eq.(%
\ref{eq176}), via regression on simple numerical values. This can be done
for a particular physical system. Here, however, we will return to the
traditional exponents in order to exhibit some associations we consider to
be general.

We can now collect the results of the analysis. In the left column we write
the approximative behavior of the contributions and in the right column the
physical origin, according to this theory. 
\begin{equation*}
\begin{array}{ll}
k^{-2} & 
\begin{array}{l}
\text{gas of vortices} \\ 
\text{(from }N=1\text{ to closely packed)}
\end{array}
\\ 
k^{-4} & 
\begin{array}{l}
\text{background turbulence + vortices }\frac{2D}{\left( k^{2}+\alpha
_{s}\right) ^{2}} \\ 
\text{weak interaction of vortices }K_{0}
\end{array}
\\ 
\sim S\left( \mathbf{k}\right) k^{-2}\;\text{(weak }\rho _{s}^{2}/A\text{)}
& \text{geometry of c.s., \emph{via} interaction} \\ 
S\left( \mathbf{k}\right) \left[ 1+f\left( \mathbf{k}\right) \right] & \text{%
perturbed c.s., \emph{via} rare events of large }N \\ 
b\left( 2D\beta \right) ^{2}\frac{1}{\left( k^{2}+\alpha _{s}\right) ^{2}}%
\frac{1}{k^{2}\left( k^{2}+\alpha _{s}\right) ^{3/2}} & \text{one-loop mode
coupling}
\end{array}
\end{equation*}

We see two ways in which these results can be useful.

First, the combination of various exponential dependences from the listed
contributions will result in an overall dependence with exponents that may
be compared with the experiments or numerical simulations \cite{KMcWT}, \cite
{KTMcWP}, \cite{MontSeyler}, \cite{Montg7}, \cite{0503424Guzdar}. The weight
of each contribution is dependent on factors like the shape of the vortices
(including the spatial extension compared to the area $A$), amplitude of the
vortices, strength of the random drive ($D$). Also in the problem enters as
parameter the function of distribution on the plane (assumed here uniform).
The energy of interaction between vortices may be reconsidered along the
example of vortices in superfluids.

Second, the spectrum can be seen as dominated in different spectral domains
by one or another of these contributions. This is compatible with the known
fact that the spectrum has regions with different exponents $\mu $.

In both these ways we have to solve an inverse problem, starting from the
experimental (or numerical) spectrum and using the above list to map the
exponential form to a physical process which may have been at its origin.

Various extensions of this treatment are possible. One can introduce a
chemical potential for the description of the statistical equilibrium
consisting of generation and suppression of coherent vortices. There are
treatments of this type for the conversion of the global rotation of
superfluids into localised vortices, an example that may also be useful for
the consideration of the zonal flow saturation in tokamak, besides the
Kelvin-Helmholtz instability and the collisions.

The treatment by generating functional allows in principle determination of
statistical correlations at any high order desired. However, while the
functional derivatives can easily produce the $n$-th order cumulant, we must
be sure that the generating functional has been calculated with the
necessary precision, or, in other words, we must be sure that we have
incorporated the physical origin of these correlations. This method should
be accompagned by the more standard analysis of closure of the hierarchy of
equations for the correlations and these two approaches must be seen as
complementary.

As a final remark, the physical model adopted in the present treatment may
be extended to cover more complex regimes, with, of course, a certain
increase in the analytical work.

\textbf{Aknowledgments}. The authors gratefully aknowledge the discussions
with Professor David Montgomery. This work has been partly supported by the
Japan Society for the Promotion of Science. The authors are grateful for
this support and for the hospitality of Professor S.-I. Itoh and Professor
M. Yagi.

\section{Appendix A : physics of the equation}

It is of interest to study a realistic two-dimensional model, like
Hasegawa-Wakatani or similar. However we need for the beginning a simpler
equation and if possible with a vortex solution with known analytical
expression.

A possibility is the equation for the ion drift instabilities but here we
must specify the scales.

\bigskip

Consider the equations for the ITG model in two-dimensions with adiabatic
electrons: 
\begin{eqnarray}
\frac{\partial n_{i}}{\partial t}+\mathbf{\nabla \cdot }\left( \mathbf{v}%
_{i}n_{i}\right) &=&0  \label{a1} \\
\frac{\partial \mathbf{v}_{i}}{\partial t}+\left( \mathbf{v}_{i}\cdot 
\mathbf{\nabla }\right) \mathbf{v}_{i} &=&\frac{e}{m_{i}}\left( -\mathbf{%
\nabla }\phi \right) +\frac{e}{m_{i}}\mathbf{v}_{i}\times \mathbf{B}  \notag
\end{eqnarray}
We assume the quasineutrality 
\begin{equation}
n_{i}\approx n_{e}  \label{a2}
\end{equation}
and the Boltzmann distribution of the electrons along the magnetic field
line 
\begin{equation}
n_{e}=n_{0}\exp \left( -\frac{\left| e\right| \phi }{T_{e}}\right)
\label{a3}
\end{equation}
In general the electron temperature can be a function of the radial variable 
\begin{equation}
T_{e}\equiv T_{e}\left( x\right)  \label{a4}
\end{equation}

The velocity of the ion fluid is perpendicular on the magnetic field and is
composed of the diamagnetic, electric and polarization drift terms 
\begin{eqnarray}
\mathbf{v}_{i} &=&\mathbf{v}_{\perp i}  \label{a5} \\
&=&\mathbf{v}_{dia,i}+\mathbf{v}_{E}+\mathbf{v}_{pol,i}  \notag \\
&=&\frac{T_{i}}{\left| e\right| B}\frac{1}{n_{i}}\frac{dn_{i}}{dr}\widehat{%
\mathbf{e}}_{y}  \notag \\
&&+\frac{-\mathbf{\nabla }\phi \times \widehat{\mathbf{n}}}{B}  \notag \\
&&-\frac{1}{B\Omega _{i}}\left( \frac{\partial }{\partial t}+\left( \mathbf{v%
}_{E}\cdot \mathbf{\nabla }_{\perp }\right) \right) \mathbf{\nabla }_{\perp
}\phi  \notag
\end{eqnarray}
The diamagnetic velocity will be neglected. Introducing this velocity into
the continuity equation, one obtains an equation for the electrostatic
potential $\phi $.

Before writing this equation we introduce new dimensional units for the
variables. 
\begin{equation}
\phi ^{phys}\rightarrow \phi ^{\prime }=\frac{\left| e\right| \phi ^{phys}}{%
T_{e}}  \label{sc1}
\end{equation}
\begin{equation}
\left( x^{phys},y^{phys}\right) \rightarrow \left( x^{\prime },y^{\prime
}\right) =\left( \frac{x^{phys}}{\rho _{s}},\frac{y^{phys}}{\rho _{s}}\right)
\label{sc2}
\end{equation}
\begin{equation}
t^{phys}\rightarrow t^{\prime }=t^{phys}\Omega _{i}  \label{sc3}
\end{equation}
The new variables $\left( t,x,y\right) $ and the function $\phi $ are
non-dimensional. In the following the \emph{primes} are not written.

\subsection{The equation}

With these variables the equation obtained is 
\begin{eqnarray}
&&\frac{\partial }{\partial t}\left( 1-\mathbf{\nabla }_{\perp }^{2}\right)
\phi  \label{a6} \\
&&-\left( -\mathbf{\nabla }_{\perp }\phi \times \widehat{\mathbf{n}}\right)
\cdot v_{d}  \notag \\
&&+\left( -\mathbf{\nabla }_{\perp }\phi \times \widehat{\mathbf{n}}\right)
\cdot \mathbf{v}_{T}\phi  \notag \\
&&+\left[ \left( -\mathbf{\nabla }_{\perp }\phi \times \widehat{\mathbf{n}}%
\right) \cdot \mathbf{\nabla }_{\perp }\right] \left( -\mathbf{\nabla }%
_{\perp }^{2}\phi \right)  \notag \\
&=&0  \notag
\end{eqnarray}
where 
\begin{eqnarray}
v_{d} &\equiv &-\mathbf{\nabla }_{\perp }\ln n_{0}-\mathbf{\nabla }_{\perp
}\ln T_{e}  \label{vstarvt} \\
\mathbf{v}_{T} &\equiv &-\mathbf{\nabla }_{\perp }\ln T_{e}  \notag
\end{eqnarray}
(This is Eq.(8) from the paper \cite{LaedkeSpatschek1}).

\subsection{No temperature gradient}

From the various versions of the nonlinear drift equation, (in particular
the Hasegawa-Mima equation ) we choose the radially symmetric
Flierl-Petviashvili soliton equation: 
\begin{equation}
\left( 1-\rho _{s}^{2}\nabla _{\perp }^{2}\right) \frac{\partial \varphi }{%
\partial t}+v_{d}\frac{\partial \varphi }{\partial y}-v_{d}\varphi \frac{%
\partial \varphi }{\partial y}=0  \label{eqh1}
\end{equation}
where $\rho _{s}=c_{s}/\Omega _{i}$, $c_{s}=\left( T_{e}/m_{i}\right) ^{1/2}$
and the potential is scaled as $\varphi =\frac{L_{n}}{L_{T_{e}}}\,\frac{%
e\Phi }{T_{e}}$ . Here $L_{n}$ and $L_{T}$ are respectively the gradient
lengths of the density and temperature. The velocity is the diamagnetic
velocity $v_{d}=\frac{\rho _{s}c_{s}}{L_{n}}$. The condition for the
validity of this equation are: $\left( k_{x}\rho _{s}\right) \left( k\rho
_{s}\right) ^{2}\ll \eta _{e}\frac{\rho _{s}}{L_{n}}$, where $\eta _{e}=%
\frac{L_{n}}{L_{T_{e}}}$.

The exact solution of the equation is 
\begin{equation}
\varphi _{s}\left( y,t;y_{0},u\right) =-3\left( \frac{u}{v_{d}}-1\right)
\sec h^{2}\left[ \frac{1}{2\rho _{s}}\left( 1-\frac{v_{d}}{u}\right)
^{1/2}\,\left( y-y_{0}-ut\right) \right]  \label{eqh7}
\end{equation}
where the velocity is restricted to the intervals $u>v_{d}$\ \ \ or\ \ \ $%
u<0 $. In the Ref. \cite{MH} the radial extension of the solution is
estimated as: $\left( \Delta x\right) ^{2}\sim \rho _{s}L_{n}$. In our work
we shall assume that $u$ is very close to $v_{d}$ , $u\gtrsim v_{d}$ (i.e.
the solitons have small amplitudes).

\section{Appendix B : Connection between the MSR formalism and
Onsager-Machlup}

In our approach the most natural way of proceeding with a stochastic
differential equation is to use the MSR type reasoning in the Jensen
formulation. The equation is discretized in space and time and selected with 
$\delta $ functions in an ensemble of functions (actually in sets of
arbitrary numbers at every point of discretization). The result is a
functional integral. There is however a particular aspect that needs careful
analysis, as mentioned in the previous Subsection. It is the problem of the
Jacobian associated with the $\delta $ functions. This problem is discussed
in Ref.\cite{DeDoPel} and for consistency we include here the essential of
their original treatment.

The equation they analyse is in the time domain and is presented in most
general form as 
\begin{equation*}
\frac{\partial \phi _{j}\left( t\right) }{\partial t}=-\left( \Gamma
_{0}\right) _{jk}\frac{\delta H}{\delta \phi _{k}\left( t\right) }+V_{j}%
\left[ \phi \left( t\right) \right] +\theta _{j}
\end{equation*}
where the number of stochastic equations is $N$ , $H$ is functional of the
fields, $V_{j}$ is the streaming term which obeys a current-conserving type
relation 
\begin{equation*}
\frac{\delta }{\delta \phi _{j}}V_{j}\left[ \phi \right] \exp \left\{ -H%
\left[ \phi \right] \right\} =0
\end{equation*}
The noise is $\theta _{j}$.

The following generating functional can be written 
\begin{equation*}
Z_{\theta }=\int \emph{D}\left[ \phi _{j}\left( t\right) \right] \exp \int dt%
\left[ l_{j}\phi _{j}\left( t\right) \right] \prod_{j,t}\delta \left( \frac{%
\partial \phi _{j}\left( t\right) }{\partial t}+K_{j}\left[ \phi \left(
t\right) \right] -\theta _{j}\right) J\left[ \phi \right]
\end{equation*}
the functions $l_{j}\left( t\right) $ are currents, 
\begin{equation*}
K_{j}\left[ \phi \left( t\right) \right] \equiv -\left( \Gamma _{0}\right)
_{jk}\frac{\delta H}{\delta \phi _{k}\left( t\right) }+V_{j}\left[ \phi
\left( t\right) \right]
\end{equation*}
and $J\left[ \phi \right] $ is the Jacobian associated to the Dirac $\delta $
functions in each point of discretization.

The Jacobian can be written 
\begin{equation*}
J=\det \left[ \left( \delta _{jk}\frac{\partial }{\partial t}+\frac{\delta
K_{j}\left[ \phi \right] }{\delta \phi _{k}}\right) \delta \left(
t-t^{\prime }\right) \right]
\end{equation*}
Up to a multiplicative constant 
\begin{equation*}
J=\exp \left( Tr\ln \left[ \left( \frac{\partial }{\partial t}+\frac{\delta K%
}{\delta \phi }\right) \frac{\delta \left( t-t^{\prime }\right) }{\frac{%
\partial }{\partial t}\delta \left( t-t^{\prime }\right) }\right] \right)
\end{equation*}
or 
\begin{equation*}
J=\exp \left( Tr\ln \left[ 1+\left( \frac{\partial }{\partial t}\right) ^{-1}%
\frac{\delta K\left( t\right) }{\delta \phi \left( t^{\prime }\right) }%
\right] \right)
\end{equation*}
Since the operator $\left( \frac{\partial }{\partial t}\right) ^{-1}$ is
retarded, only the lowest order term survives after taking the trace 
\begin{equation*}
J=\exp \left[ -\frac{1}{2}\int dt\frac{\delta K_{j}\left[ \phi \left(
t\right) \right] }{\delta \phi _{j}\left( t\right) }\right]
\end{equation*}
The factor $1/2$ comes from value of the $\Theta $ function at zero.

In the treatment which preserves the dual function $\widehat{\phi }$
associated to $\phi $ in the functional, there is a part of the action 
\begin{equation*}
\widehat{\phi }K\left[ \phi \right]
\end{equation*}
Then a $\widehat{\phi }$ and a $\phi $ of the same coupling term from $%
\widehat{\phi }K\left[ \phi \right] $ close onto a loop. Since $G_{\widehat{%
\phi }\phi }$ is retarded, all these contributions vanish except the one
with a single propagator line. This cancels exactly, in all orders, the part
coming from the Jacobian.

Then it is used to ignore all such loops and together with the Jacobian.

In conclusion we can compared the two starting points in a functional
approach: The one that uses \emph{dual functions} $\phi \left( t\right) $
and $\chi \left( t\right) $, closer in spirit to MSR; And the approach based
on Onsager-Machlup functional, traditionally employed for the determination
of the probabilities. Either we keep $\chi \left( t\right) $ and ignore the
Jacobian (the first approach) or integrate from the beginning over $\chi
\left( t\right) $ and include the Jacobian. The approaches are equivalent
but we have followed the first one.

\end{document}